\documentclass[iop,useAMS]{emulateapj}
\usepackage{amsmath}
\usepackage[colorlinks, citecolor=blue]{hyperref}

\shorttitle{The mass and radius of Aql X--1}
\shortauthors{Z.S. Li et al.}

\begin{document}
	
\title{Simultaneous constraints on the mass and radius of Aql X--1 from quiescence and X-ray burst observations}
	
\author{Zhaosheng Li \altaffilmark{1,2,3}}
\email{lizhaosheng@xtu.edu.cn}
	
\author{Maurizio Falanga \altaffilmark{3}}
	
\author{Li Chen \altaffilmark{4}}
	
\author{Jinlu Qu \altaffilmark{5}}
	
\author{Renxin Xu \altaffilmark{6,7}}

\altaffiltext{1}{Department of Physics, Xiangtan University, Xiangtan, 411105, P.R. China}
\altaffiltext{2}{International Space Science Institute, Hallerstrasse 6, 3012 Bern, Switzerland}
\altaffiltext{3}{Albert Einstein Center for Fundamental Physics, Institute for Theoretical Physics / Laboratory for High-Energy Physics, University of Bern }
\altaffiltext{4}{Department of Astronomy, Beijing Normal University, Beijing 100875, China}
\altaffiltext{5}{Laboratory for Particle Astrophysics, Institute of High Energy Physics, CAS, Beijing 100049, P.R. China}
\altaffiltext{6}{Kavli Institute for Astronomy and Astrophysics, Peking University, Beijing 100871, P. R. China  (FAST Fellow distinguished)}
\altaffiltext{7}{School of Physics and State Key Laboratory of Nuclear Physics and Technology, Peking University, Beijing 100871, P.R. China}
	
\begin{abstract}
The measurement of neutron star mass and radius is one of the most direct way to distinguish between various dense matter equations of state. The mass and radius of accreting neutron stars hosted in low mass X-ray binaries can be constrained by several methods, including photospheric radius expansion from type-I X-ray bursts and from quiescent spectra. In this paper, we apply for the first time these two methods simultaneously to constrain the mass and radius of \object{Aql X--1}, as a reliable  distance estimation, high signal-to-noise ratio quiescent spectra from \textit{Chandra} and \textit{XMM-Newton}, and photospheric radius expansion bursts from \textit{RXTE} are available. This is also used to verify the consistency between the two methods, and to narrow down the uncertainties of the neutron star mass and radius. It is found that the distance to Aql X--1 should be in the range of $4.0-5.75$ kpc, based on the overlapping confidence regions between photospheric radius expansion burst and quiescent spectra methods. In addition, we show that the mass and radius determined for the compact star in \object{Aql X--1} are compatible with strange star equations of state and conventional neutron star models.
		
\end{abstract}


\section{Introduction}
\label{sec:intro}
Accurate measurements of neutron star (NS) masses and radii provide the tightest constraints on the equations of state (EoSs) of these objects, i.e., the relation between the pressure and the supra-nuclear density in their interiors \citep{Lattimer12}. NS masses can be determined in low or high-mass X-ray binary systems hosting a main sequence companion star, or with high accuracy in double NSs or NS--white dwarf systems, through  their mass functions or pulse arrival time, respectively \citep[see e.g.,][]{Taylor92, Weisberg10,Watts15}. However, the precise and contemporaneous measurements of NSs masses and radii (mass-radius ratio) are still challenging  \citep[see e.g.,][and references therein]{Miller16}.
	
Several methods have been proposed to constrain NS masses and radii, e.g., using type-I X-ray bursts exhibiting photospheric radius expansion (PRE; see Sec. \ref{sec:burst})  \citep{Sztajno87}, modeling thermal emission from quiescent low-mass X-ray binaries (LMXBs; see Sec. \ref{sec:quie}) \citep{Heinke06}, modeling the X-ray pulse profile of accreting millisecond X-ray pulsars \citep[see e.g.,][]{Poutanen03}, or measuring the gravitational redshift of spectral features produced in the NS photosphere \citep{Ozel06}. The gravitational redshift provides one of the most accurate and model independent method to obtain the NS mass-radius ratio. Currently, a debated measurement of a large gravitational redshift measurement, $z=0.35$, was claimed for absorption lines in the X-ray burst spectra of the NS EXO 0748-676 \citep{Cottam02}, but the results were not confirmed in subsequent publications \citep{Sidoli05, Cottam08}.

LMXB systems, hosting a X-ray pulsar, accretes matter via Roche-lobe overflow from a main-sequence donor star ($M < 1 M_{\odot}$), forming  a disk around the compact object. The accreted matter is channeled by the magnetic field lines toward the magnetic poles which produce by impact two hot spots on the NS surface. Therefore, when the NS spins, the observed  X-ray pulses are modulated by relativistic effects (e.g., light bending, Doppler boosting, and aberration), which depends on the compactness term $M/R$ \citep{Pechenick83}. Modeling the pulse profiles from X-ray pulsars can provide strong constraints on the NS mass and radius, although some degeneracies due to unknown factors like the geometry of the hot spot and the observer inclination have to be taken account \citep[see e.g.,][]{Beloborodov02,Poutanen03, Leahy04,Bogdanov07,Leahy09}.
To break the degeneracies among various parameters and to quantify the mass and radius simultaneously, \citet{Psaltis14} utilized the properties of the fundamental and the second harmonic of the pulse profile. However, a larger number of processes should be considered, such as the oblateness of the NS surface, the quadrupole moment, the pulse profile variation, the geometrical factors, and the NS atmosphere emission \citep{Morsink07,Baubock15,Hartman08,Psaltis14a, Psaltis14}. Considering these effects, the Neutron star Interior Composition ExploreR mission (NICER),  expected to be launched mid-2017, is designed to determine at 10\% of accuracy the NS radius from X-ray pulsars  \citep{Gendreau12}. If a pulse profile of an X-ray pulsar shows long time evolution, implying that the hot spots are drifting on the NS surface, then large area detectors are needed to collect enough photons on short time scales \citep{Zhang16}. 	
	
\subsection{NS mass-radius relation from type-I X-ray bursts with PRE}
\label{sec:burst}

The accreted matter on a NS surface can trigger hydrogen/helium or mixed thermonuclear flashes, called type-I X-ray bursts  \citep[see e.g.,][]{Lewin93}. The total burst energy released are on the order of $\sim 10^{39-42}$ erg, and the spectra are described by a blackbody with a temperature,  $kT_{\rm bb}$, and its normalization, $K$ \citep[e.g.,][]{Galloway08}. The energy-dependent decay time of these bursts is attributed to the cooling of the NS photosphere and results in a gradual softening of the burst spectrum \citep[see][for a review]{Lewin93,Strohmayer06}. During some type I X-ray bursts, the energy release is high enough that the luminosity reaches the Eddington limit, $L_{\rm Edd}\sim 2\times10^{38}$ erg s$^{-1}$, i.e., the value at which the gravity balances the radiative pressure. At that luminosity level, the radiation pressure lifts the surface layers from the NS in a PRE episode. Since the luminosity scales as $L_{\rm burst} \propto R_{\rm bb}^2 (kT_{\rm bb})^4$ for pure blackbody spectrum, during the PRE episode, while the bolometric luminosity remains constant at the Eddington value, the temperature, $kT_{\rm bb}$, drops when the radius of the photosphere, $R_{\rm bb}$, expands. The point at which the NS atmosphere reaches the surface again, $R_{\rm bb,~min} = R_{\rm NS}$, i.e., at the highest temperatures, is called {\it touchdown}. To derive the bolometric burst flux a color correction factor, $f_{\rm c}=T_{\rm bb}/T_{\rm eff}$ should be applied, since the burst emitted photons are up-scattered in a hot NS atmosphere \citep{Ebisuzaki84}.  

 Assuming spherically symmetric emission from a non-spinning NS surface, the Eddington luminosity is expressed as \citep{Lewin93}:
\begin{equation}
L_{\rm{Edd}}=4\pi D^2F_{\rm TD}=\frac{4\pi GMc}{\kappa_{\rm es}} \left(1-\frac{2GM}{Rc^2}\right)^{1/2},
\label{eq:LEdd}
\end{equation}
where, $G$, is the gravitational constant,  $c$, the speed of light, $D$ the distance to the source, $M$ and $R$  the NS mass and radius, respectively. The parameter $\kappa_{\rm es}=0.2(1+X)$ is the electron scattering opacity, $X$ the atmosphere's hydrogen mass fraction ($X=1$ is for pure hydrogen), and $F_{\rm TD}$ the Eddington flux at the touchdown. After the touchdown moment, the NS photosphere cools down on the whole surface, while the emission area remains nearly constant. This has been observed in a substantial fraction of X-ray bursts \citep{Guver12a}. Given the distance to the source, the blackbody normalization can be written as  \citep{Ozel06}:
\begin{equation}
K=\frac{1}{f^4_{\rm c}}\frac{[R(1+z)]^2}{D^2}=(\,f_{\rm c}A)\,^{-4}, 
\label{equ:area}
\end{equation} 
where $A=(R(1+z)/D)^{-1/2}$ is the apparent angular size, $1+z=(1-2GM/Rc^2)^{-1/2}$, and $z$ is the gravitational redshift. Therefore, the NS mass and radius can be constrained from the observed $K$ and $F_{\rm TD}$, once the atmosphere's hydrogen mass fraction, $X$, the color correction factor, $f_{\rm c}$, and the source distance are known. We note, that this ``touchdown" method assumes that the color correction factor is constant during the whole burst cooling phase. However, based on NS hot atmosphere  models, while the emission approch the Eddington limit, the color correction factor rises rapidly \citep{Suleimanov11, Suleimanov12}. In this case, the $F-K^{-1/4}$ track denotes the dependence between $f_{\rm c}$ and the flux, afterwards, the $F_{\rm Edd}$ and $A$ are fitted from the $F-K^{-1/4}$ curve to estimate the NS mass and radius. That is the so called ``cooling tail" method. A study of \object{4U 1608-52} showed that  the $F-K^{-1/4}$ curves follow the theoretical  atmosphere model only if the source is in a low accretion (or hard) state \citep{Poutanen14}. Therefore, to constrain the mass and radius following the predicted cooling tail shape, the authors of this work proposed to study only PRE bursts occurring during a hard spectral state \citep{Suleimanov11, Suleimanov12}. 
This is in contradiction to the results published by \citet{Ozel15a} that the mass and radius of \object{4U 1608--52} are best determined using bursts with the brightest touchdown. It is challenging, for some bursts, to follow the very step cooling tail after touchdown, as the limited time resolution of \textit{RXTE} could not resolve the blackbody normalization evolution at these phases \citep{Ozel15a}.

For a NS atmosphere model, considering the energy dependent Klein-Nishina cross section, the  Eddington luminosity, Eq. (\ref{eq:LEdd}), can be rewritten as \citep[see][and references therein]{Ozel16}:

\begin{eqnarray}
 F_{\rm TD}=&&\frac{GMc}{k_{\rm es}D^2}\left(1-\frac{2GM}{Rc^2}\right)^{1/2} \times \nonumber\\ 
 &&\left[1+\left(\frac{kT_{\rm TD}}{38.8~{\rm keV}}\right)^{a_g}\left(1-\frac{2GM}{Rc^2}\right)^{-a_g/2}\right],\label{equ:flux_cor}
\end{eqnarray}
where $T_{\rm TD}$ is the critical temperature at the Eddington limit during  touchdown, and $a_g=1.01+0.067({g_{\rm eff}}/{10^{14}~{\rm cm~ s^{-2}}})$ with $g_{\rm eff}=GMR^{-2}(1-2GM/Rc^2)^{-1/2}$  the effective gravitational acceleration constant. 
The blackbody normalization, Eq. (\ref{equ:area}), can be rewritten considering the Doppler spectrum broadening, the oblateness, and  the NS quadrupole moment with spin frequencies larger than $\nu_{\rm NS}=400~{\rm Hz}$  \citep{Baubock15, Ozel16}:
\begin{equation}\label{equ:area_cor}
\begin{split}
 &K = \frac{R^2}{D^2 f_{\rm c}^4}\left(1-\frac{2GM}{Rc^2}\right)^{-1} \left\{1+\left[ {\left(0.108-0.096\frac{M}{M_\odot}\right) } \right.\right.-  \\ 
& \left.\left. {\left(0.061-0.114\frac{M}{M_\odot}\right)\frac{R}{10~{\rm km}} -0.128\left(\frac{R}{10~{\rm km}}\right)^2}\right]\left(\frac{\nu_{\rm NS}}{10^3~{\rm Hz}}\right)^2\right\}^2. 
 \end{split}
\end{equation}

Given the large number of parameters to determine the NS mass and radius, two different  distribution probability
approaches have been introduced, i.e., the Bayesian and frequentist  \citep{Ozel09, Ozel15a}. For a given pair of NS mass and radius $(M, R)$, the equation of the Bayesian probability to obtain the observed touchdown flux, $F_{\rm TD}$, and blackbody normalization, $K$, is:

\begin{equation}\label{equ:prob}
\begin{split}
 &P({\rm data|M,R})=\int P(D){\rm d}D\int P(f_{\rm c}){\rm d}f_{\rm c} \int P(kT_{\rm TD}){\rm d}kT_{\rm TD}\times \\
 &\int P(X){\rm d}X  P[F_{\rm TD}(M, R, D, X, kT_{\rm TD})]P[K(M, R, D, f_{\rm c})]. 
 \end{split}
\end{equation}

The frequentist approach differs from the Bayesian approach by an additional term, i.e., the Jacobian factor $J=2cGR|1-4GM/Rc^2|/D^2f^2_ck_{\rm es}(1-2GM/Rc^2)^{3/2}$ \citep{Ozel15a}. The Jacobian factor is equal to zero for $R=4GM/c^2$ in the $M-R$ plane, leading to two separated $M-R$ solutions. If the inferred $M-R$ values are close to the $R=4GM/c^2$ separation line, the NS $M-R$ are biased. At variance, for the Baysian approach method, if the M/R values approach the $R=4GM/c^2$ line, one solution exists. We thus choose to constrain the mass and radius of Aql X--1 adopting the Baysian approach \citep{Ozel15a, Ozel16}.

For the touchdown method, we constrain the NS $M-R$ values using Eqs. (\ref{equ:flux_cor}), (\ref{equ:area_cor}), (\ref{equ:prob}), and by taking into account the source distance, atmosphere composition, touchdown flux and temperature, blackbody normalization, and color correction factor uncertainties. Following \citet{Ozel16}, we consider  blackbody normalization values in the range $(0.1-0.7)~F_{\rm TD}$.  The upper value is chosen to overcome the potential missing touchdown moment, and the lower value 
is set to avoid a partial emission, occurring in case the cooling luminosity becomes so faint that may lead to an underestimated  blackbody normalization.
Moreover, for the selected blackbody normalization range, $f_c$, is nearly invariant for several hot atmosphere models with different metal abundances and surface gravities \citep[see e.g.,][]{Madej04,Majczyna05,Suleimanov11,Suleimanov12}.

\subsection{NS mass-radius relation from quiescent spectra}
\label{sec:quie}

The spectrum of quiescent low-mass X-ray binaries (qLMXBs) can be modeled by two-components: a soft thermal component together with a hard non-thermal component  \citep{Rutledge01}. We note that for most of the qLMXBs in globular cluster the emission spectrum is dominated by  a soft thermal emission with a weak or non-existent non-thermal component \citep{Guillot13}. The soft thermal component is interpreted as being due to deep crustal heating, mainly emitted thought the NS atmosphere \citep{Brown98}. NS masses and radii can thus be predicted by fitting the qLMXBs spectra with atmosphere models, e.g. {\sc nsatmos} \citep{Heinke06}, {\sc nsa} \citep{Zavlin96}, or {\sc mcphac} \citep{Haakonsen12} \citep{Heinke06, Webb07, Guillot11, Guillot13, Guillot14, Bogdanov16}. In all these models a pure hydrogen atmosphere is assumed unless a hydrogen deficit companion star is observed, by, e.g., measuring an upper limit of the H$\alpha$-line equivalent width \citep{Heinke14}.

The knowledge of the hard spectrum is critical, since the hard component may affect the soft spectrum during the simultaneous fit.
The hard component is usually interpreted as being due to residual accretion \citep{Rutledge01, Campana04, Cackett10, Bernardini13, Chakrabarty14}, or emission from either boundary layer (where the accretion flow hits the NS surface) or a radiatively-inefficient accretion flow model \citep{Chakrabarty14, Angelo15}.

 For \object{Cen X-4}, X-ray flares have been observed during the quiescent state, where the hard X-ray photons ($2-10$ keV) are positively correlated with the soft X-ray photons ($0.3-2.0$ keV). The total X-ray spectrum is correlated with the ultraviolet and optical disk emission \citep{Bernardini13,Cackett10,Coti14}. Therefore, the total X-ray emission from \object{Cen X-4}  may arise solely from a continuous and variable low level accretion flow \citep[e.g.,][]{Zampieri95,Zampieri01}. This interpretation excludes soft X-ray NS atmosphere models, and thus, no constraints on mass and radius can be derived. However, for \object{Aql X--1} during flares in the quiescent state, no correlation between the hard and soft components was found \citep{Cackett10,Coti14}. From the two component spectral fit models, the non-thermal component can be attributed to the X-ray variability, even thought the thermal component variation can not be completely ruled out. During the \object{Aql X--1}  outburst in 2010, the interaction between the magnetic field and the stellar rotation may explain the fast decay time back to quiescence. The post outburst quiescent X-ray luminosity correlates with the strength of the non-thermal emission, due to residual accretion \citep{Campana14}. The post outburst quiescent states have been observed also with \textit{Swift} in 2012, 2013, and 2015 \citep{Waterhouse16}. In these cases, the spectrum was fitted with the addition of a significant hard spectral component in 2012, and with pure thermal emission in 2013 and 2015 \citep{Waterhouse16}. Interestingly, even if during the 2012 and 2013 outbursts the spectral components were different, the soft component (i.e., NS temperature) evolved from the outburst to the quiescent states with the same trend. It can be summarized that for \object{Aql X--1} the thermal component is deep crustal cooling. This conclusion should be confirmed by future observations because of the low signal-to-noise ratio of the \textit{Swift} spectra.  


%
To infer the NS mass and radius, high signal-to-noise ratio \textit{Chandra} and \textit{XMM-Newton} data have been obtained for six NSs qLMXB, located in the globular clusters \object{M13}, \object{M28}, \object{M30}, \object{$\omega$ Cen}, \object{NGC 6397}, and \object{NGC 6304} \citep{Guillot13}. The soft X-ray spectra were fitted with a pure hydrogen atmosphere model, {\sc nsatmos}. The NSs spectra have been fitted simultaneous, assuming one radius value for all NSs. The determined radius is $9.1^{+1.3}_{-1.5}~{\rm km}$ \citep{Guillot13}. This value has been marginally improved to $9.4\pm1.2~{\rm km}$ by increasing the exposure of the \object{M30} and \object{$\omega$ Cen}  \textit{Chandra} observations \citep{Guillot14}, and increased to $10.3^{+1.2}_{-1.1}~{\rm km}$ by using updated distance measurements to globular clusters \citep{Guillot16}. Combining the sample of these six qLMXB sources, with the addition of six LMXBs exhibiting PRE bursts, and assuming a NS mass of $1.5~M_{\rm \odot}$, a radius of $10.8^{+0.5}_{-0.4}~{\rm km}$ is found \citep{Ozel16}. These works exclude two EoSs for the matter above the nuclear saturation density,  MS0 \citep{MS96} and PAL \citep{PAL88}  at the 99\% confidence level (c.l.) \citep{Guillot13, Guillot14} and prefer the EoS of AP4 \citep{AP97}, especially for low mass NSs \citep{Ozel16}. We note that  the observed high NS masses up to $\sim2M_{\odot}$, as well as low NS masses down to $\sim1M_{\odot}$, may constrain the EoSs too \citep{Demorest10, Antoniadis13, Li15, Falanga15}. 	



\subsection{The source Aql X--1}
\label{sec:source}

\object{Aql~X--1} is a transient LMXB, hosting a NS orbiting around a main sequence K4 spectral type companion  \citep{Callanan99, Chevalier99, Mata16}.

In the last few decades, the source was monitored with many X-ray instruments  \citep[e.g.,][]{Campana98, Campana13}. \object{Aql X--1} is know to host a fast spinning pulsar $\sim550.27~ {\rm Hz}$ with an orbital period of $\sim18.97$ hr, exhibiting frequent outbursts and type-I X-ray bursts \citep{Casella08,Campana98, Campana13,Galloway08}. The source distance has been estimated in the range of $4-6.5 ~{\rm kpc}$, based on the companion star filling its Roche Lobe to produce the frequently observed outbursts \citep{Rutledge01b}.  Recently, it has been revised to $6\pm2~{\rm kpc}$, and the uncertainty mainly comes from our limited knowledge on the  companion star type and radius \citep[see][for more details]{Mata16}. In our simulations,  we use a distance in the range of $4-6.5 ~{\rm kpc}$ (see Sec~\ref{sec:mass-radius}). 
 
From high resolution spectroscopy, obtained with \textit{XMM-Newton/RGS}, an hydrogen column density of $N_{\rm H}=(5.21\pm0.05)\times 10^{21}~{\rm cm^{-2}}$ has been measured by \citet{Pinto13}. These authors measured nitrogen K edge, iron L2 and  L3 edge, and oxygen K edge, which lead to a model independent estimate of the hydrogen column density. This parameter is fundamental to determine the NS mass and radius from its quiescent spectra (see Sec. \ref{sec:qspec}), which is applied to the case of  qLMXBs located in globular clusters. This is because  the distance to globular clusters is accurately known through optical observations, and the non-thermal emission in these objects contributes negligibly to the total flux, limiting the impact on the  determination of the NS mass and radius.  In the case of \object{Aql X--1} the non-thermal component contributes by $\geqslant$15\% to the total source flux \citep{Campana14, Coti14}. \citet{Coti14} compared the contributions of the power-law component in the quiescent spectra of the source and concluded that it is due to residual accretion heating the NS atmosphere. So, the variations of the thermal temperature and the PL normalization is enough to account for the residual accretion.

The qLMXBs hosted in globular clusters are observed to emit almost a pure thermal emission (non-thermal component flux less than 5\%), being good study cases to measure the mass and radius  \citep{Guillot13, Heinke14, Bogdanov16}. However, open questions should be still better explored in the quiescent spectral method, such as the atmosphere composition of NSs and the origin of non-thermal emission.
 Both the PRE burst and qLMXBs methods have few undetermined parameters, such as a color correction factor, or the atmosphere composition. Moreover, at present, no LMXB  in a globular cluster has high signal-to-noise ratio spectra and PRE bursts simultaneously. In this paper, we analyse \object{Aql X--1}, which has distance estimation, PRE bursts from \textit{RXTE}, and high signal-to-noise ratio spectra from \textit{XMM-Newton} and \textit{Chandra}. We constrain the mass and radius of  \object{Aql X--1} for the first time with the above mentioned methods and investigate the consistence between them.

In Sec. \ref{sec:data} we report the \object{Aql X--1} data from \textit{Chandra}, \textit{XMM-Newton}, and \textit{RXTE} observations. In Sec. \ref{sec:results}, we show the time-resolved PRE bursts and quiescent spectra results.  The derived mass and radius are reported in Sec. \ref{sec:mass-radius} and its uncertainties are discussed in Sec. \ref{sec:discussions}. Finally, we draw our conclusions on the determined \object{Aql X--1} mass and radius in Sec. \ref{sec:summary}. 
	
\section{Observations and data}
\label{sec:data}

We analysed  the quiescent spectra obtained by \textit{Chandra} and \textit{XMM-Newton} observations. The X-ray bursts were studied by exploiting \textit{RXTE} data. Our total data-set includes 15 quiescent \textit{Chandra} observations, 5 quiescent  \textit{XMM-Newton} observations, and 14 \textit{RXTE} pointings during which several PRE bursts have been detected. The \textit{Chandra} and \textit{XMM-Newton} data are obtained from previous studies \citep{Rutledge01,Cackett11,Campana14}. The log of all used observations are shown in Table \ref{table:quiescent} and \ref{table:burst}. 

\begin{deluxetable}{cccc}\centering
\tabletypesize{\scriptsize}
\tablecaption{Quiescent observations of Aql X--1.\label{table:quiescent}}
\tablewidth{0pt}
\tablecolumns{4}
\tablehead{\colhead{Obs\_ID} &\colhead{Exposure Time }  &\colhead{Detector}    & \colhead{Net Count Rate}  \\
 \colhead{{\em XMM-Newton}} &\colhead{ks}  &\colhead{}    & \colhead{cts/s  (0.5-10 keV)} 
				}			        
		\startdata
			\hline\\        
        0085180401 &7.20 &MOS1 &$0.096  \pm 0.004$   \\
                   &  &MOS2 &$0.100 \pm 0.004$   \\
                   &4.91 &PN   &$0.331 \pm 0.009$   \\        
        0085180501 & 14.42  & MOS1&$ 0.093 \pm 0.003$   \\	
                   &   & MOS2&$0.100 \pm 0.003$  \\	
                   & 11.31 & PN&$0.309 \pm 0.006$   \\	            
        0112440101 &2.46 &MOS1 &$0.058 \pm 0.005 $  \\
                   &  &MOS2 &$0.059 \pm 0.005$   \\
        0112440301 &7.08& MOS1 &$0.049 \pm  0.003$  \\
                   &  & MOS2 &$0.053 \pm 0.003$   \\
        0112440401 &13.40  &MOS1 & $0.047 \pm 0.002 $   \\
                   &  &MOS2 & $0.048 \pm 0.002 $  \\
		\hline\\
		{\em Chandra} &&&\\
		\hline\\
		708  &6.63 &ACIS-S & $ 0.182 \pm 0.005$   \\
		709  &7.79 &ACIS-S & $ 0.092 \pm 0.003$   \\
		710  &7.39 &ACIS-S & $ 0.126 \pm 0.004$  \\
		711  &9.25 &ACIS-S & $0.123 \pm 0.004$  \\
		3484  &6.60 &ACIS-S &$0.162 \pm 0.005$   \\
		3485  &6.96 &ACIS-S &$ 0.182 \pm 0.005$  \\
		3486  &6.49 &ACIS-S &$0.343 \pm 0.007$   \\
		3487  &5.94 &ACIS-S &$0.094 \pm 0.004$   \\
		3488  &6.51 &ACIS-S &$ 0.087 \pm0.004$   \\
		3489  &7.13 &ACIS-S &$ 0.079 \pm 0.003$   \\
		3490  &6.94 &ACIS-S &$0.104 \pm0.004$   \\
		7629  &9.87 &ACIS-S &$0.092\pm0.003$    \\
		12457  &6.36 &ACIS-S &$ 0.258\pm0.006$   \\
		12458  &6.36 &ACIS-S &$0.147 \pm0.005$   \\
		12459  &6.36 &ACIS-S &$0.158 \pm 0.005$  \\
\enddata
\end{deluxetable}
	
\subsection{Chandra}
\label{sec:chandra}

 The \textit{Chandra} \citep{Weisskopf00} data reduction was performed by using CIAO v4.6 with standard procedures. We reprocess the data {\sc{chandra\_repro}}\footnote{http://cxc.harvard.edu/ciao/threads/data.html} indicated in Table~\ref{table:quiescent}
to create level-2 event files. The {\sc specextract} tool was used to extract the source spectra from a circled region centered on the object with a radius of 4\arcsec (8 pixels); the background spectra were carefully extracted from a nearby source-free region with a radius of 10\arcsec (20 pixels) located on the same CCD. The observation ID. 12456 was heavily piled-up, and thus we excluded these data in our analysis \citep[see][for more details]{Campana14}. The {\sc mkacisrmf} and {\sc mkarf} scripts were used to generate the response matrix (rmf) and ancillary response files (arf). To account for the absolute flux calibration, we added 3\% systematic uncertainties in the energy range 0.5-10 keV. 

\subsection{XMM-Newton}
\label{sec:xmm}
We made use only of \textit{XMM-Newton}  \citep{Jansen01} data in image mode in order to exploit the high resolution energy spectra. The data reduction was carried out by using the {\tt emchain} and {\tt epchain} tools included in the \textit{XMM-Newton Science Analysis System}. Following the standard data reduction threads\footnote{http://www.cosmos.esa.int/web/xmm-newton/sas-threads}, we filtered the EPIC data with the FLAG==0 option and retained all events with pattern $0-4$ ($0-12$) for the pn (MOS) detector(s). The source spectra were extracted from a circle with a radius of 32\arcsec and 28\arcsec for the pn and MOS, respectively. The background pn and MOS spectra were extracted from a circular source-free region with a radius of 100\arcsec  (the background and source extraction region were located in the same CCD).  We used the  {\sc rmfgen} and {\sc arfgen} to generate the response matrices file and the ancillary response file for each observation. We added a systematic uncertainty of 3\% in the energy range 0.5-10 keV in order to take into account the instrument absolute flux calibrations. 




	
\subsection{RXTE}
\label{sec:bursts}

\textit{RXTE}  \citep{Jahoda96} observed over 1000 X-ray bursts from hundreds of LMXBs during its scientific operations \citep{Galloway08}. Thanks to its large collecting area and relative wide energy band coverage, \textit{RXTE} is well suited to carry out a time resolved spectral analysis of the X-ray bursts. We found 14 PRE bursts from \object{Aql X--1} in the type-I X-ray burst catalogue published by \citet{Galloway08} and the later works by \citet{Chen13} and \citet{Kajava14}. We note, that only one PRE burst (No. 11) occurred while the source was in a hard state based on its color color diagram position \citep{Chen13, Kajava14}. The detailed \textit{RXTE} observations are reported in Table~\ref{table:burst}. 

The burst time resolved spectra were extracted from the cleaned science event files and dead time effects were corrected following standard procedures. For each PRE burst, a 16~s spectrum prior to the burst  was extracted and used as background (thus including the contribution from the source persistent emission, the diffuse X-ray background, and the instrument background). The exposure of the time resolved spectra (3--22 keV) during each of the burst was chosen in such a way that each spectrum was characterized by a similar signal-to-noise ratio. In particular, we used an exposure time of 2~s when the source count-rate was $\leq 1500$ cts/s, 1~s for count-rates of $\leq 3000$ cts/s, 0.5~s for count-rates of $\leq 6000$ cts/s, and count-rates of 0.25~s for count-rates of $> 6000$ cts/s. We rebinned all spectra in order to have at least 20 photons in each energy bin (see Sec. \ref{sec:pre}).  A systematic error of 0.5\% was applied to \textit{RXTE}/PCA spectra which correspond to the uncertainty in the response matrix \citep{Shaposhnikov12}. The uncertainties in the burst spectral parameters are given at  $1\sigma$ c.l. for a single parameter. 

\begin{figure*}\centering
\plottwo{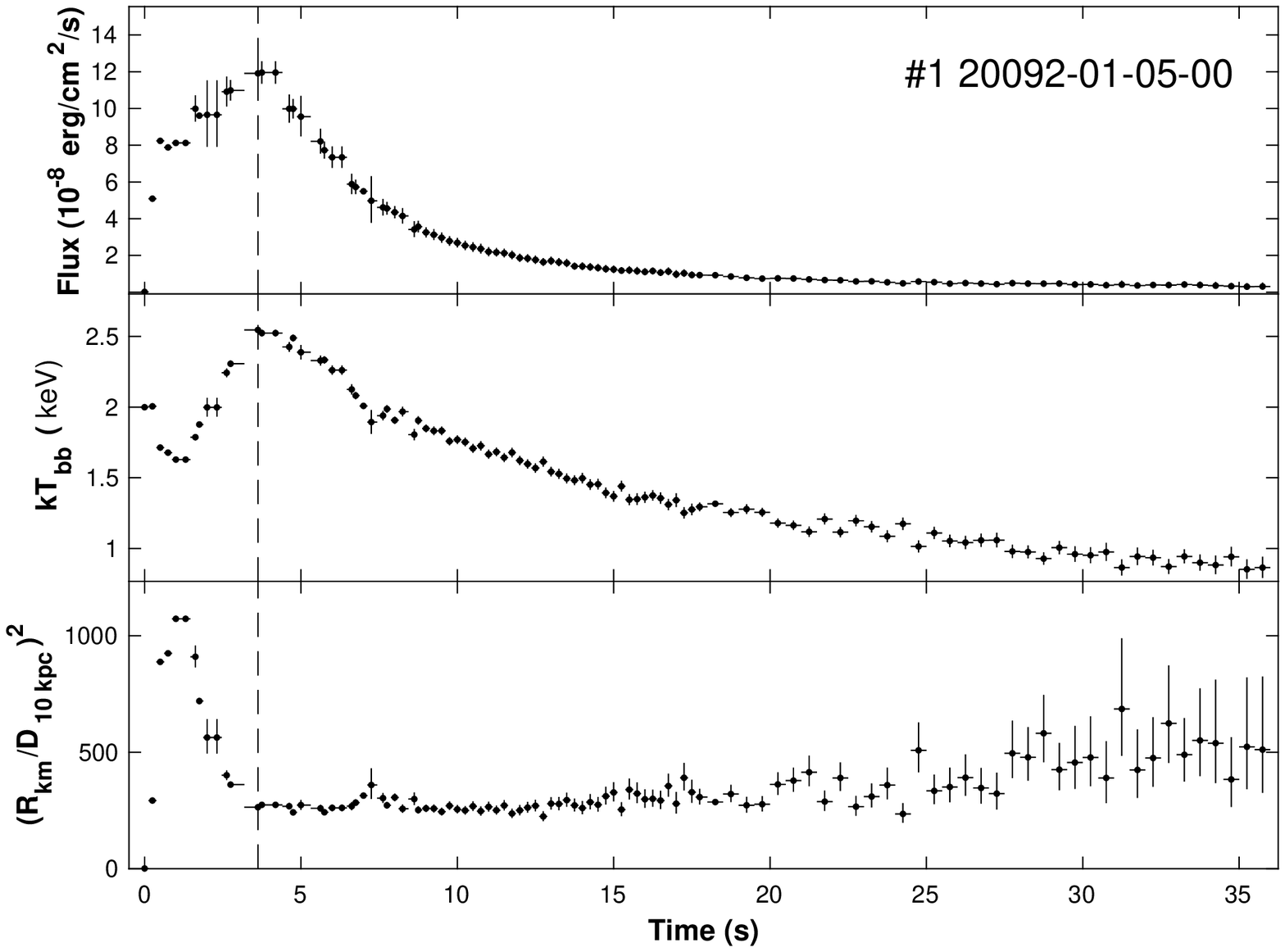}{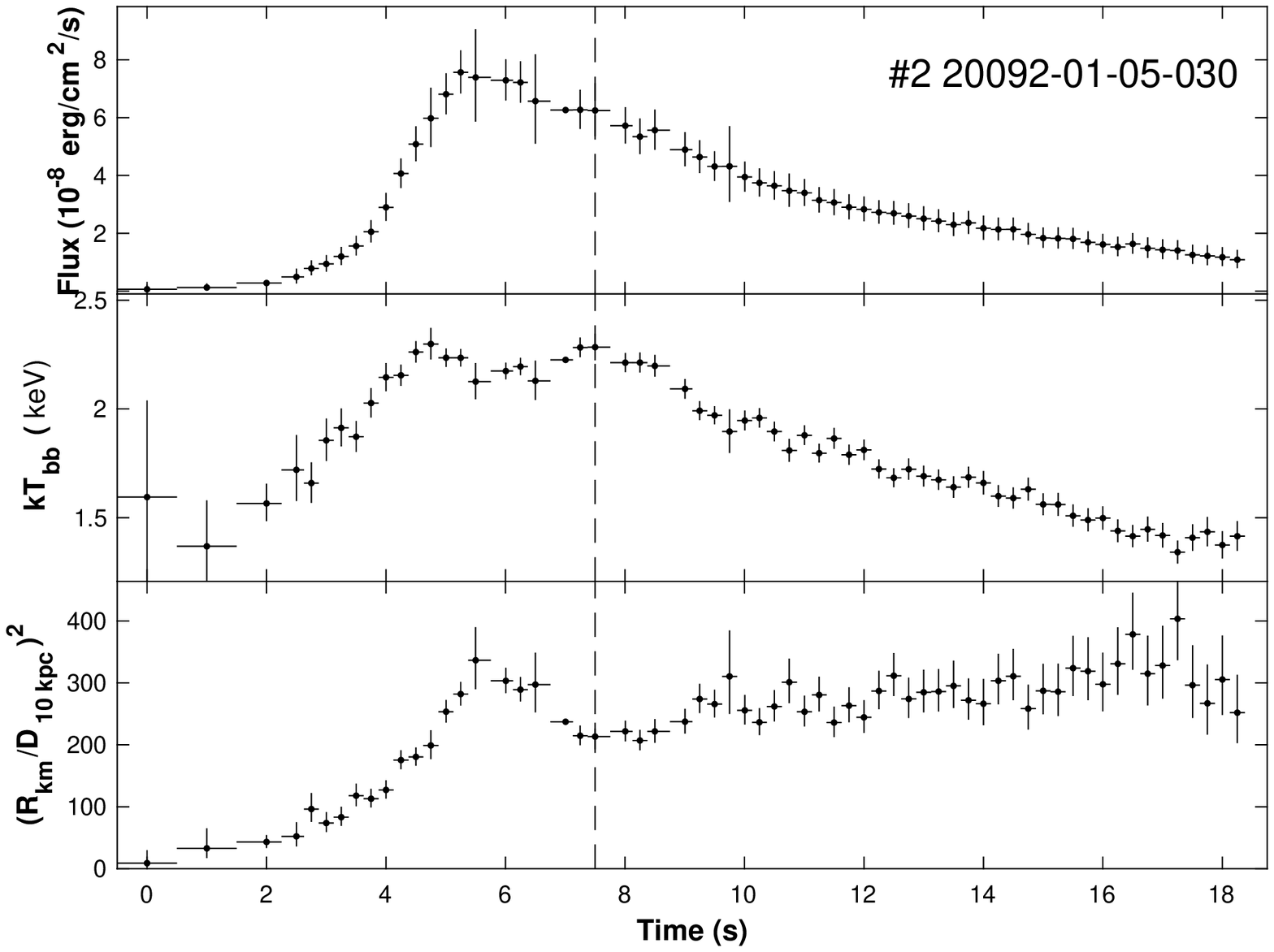}
\plottwo{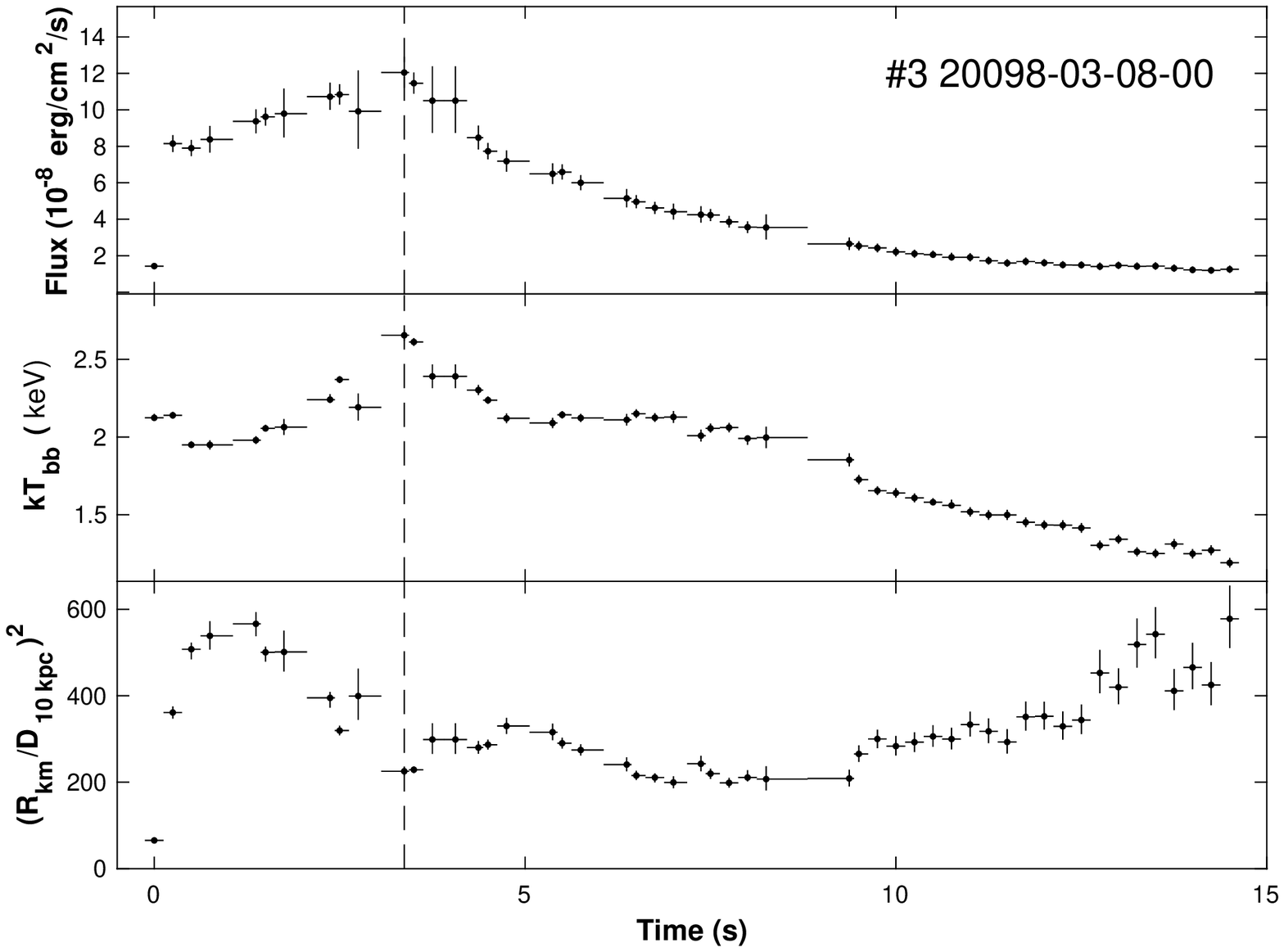}{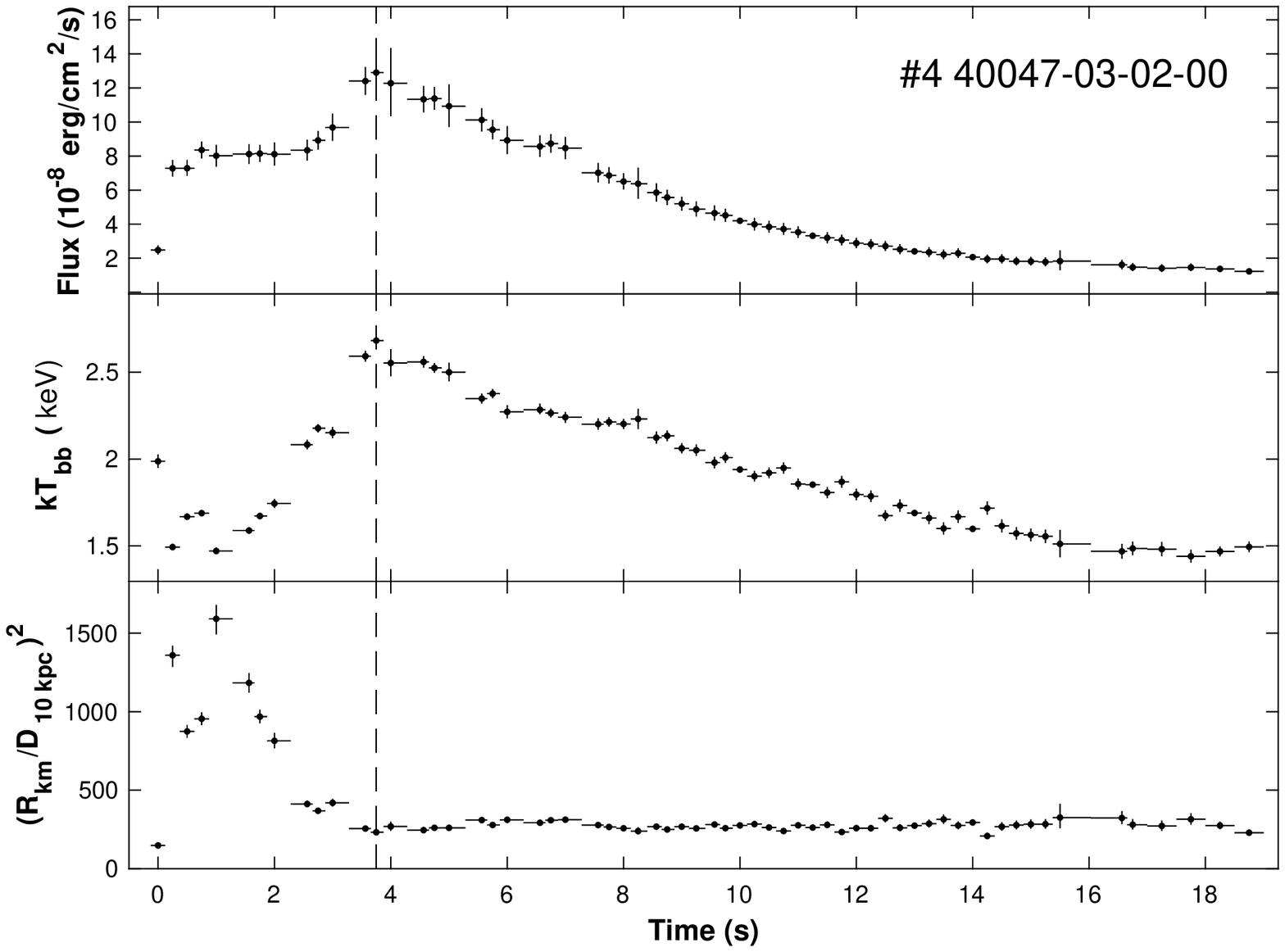}
\plottwo{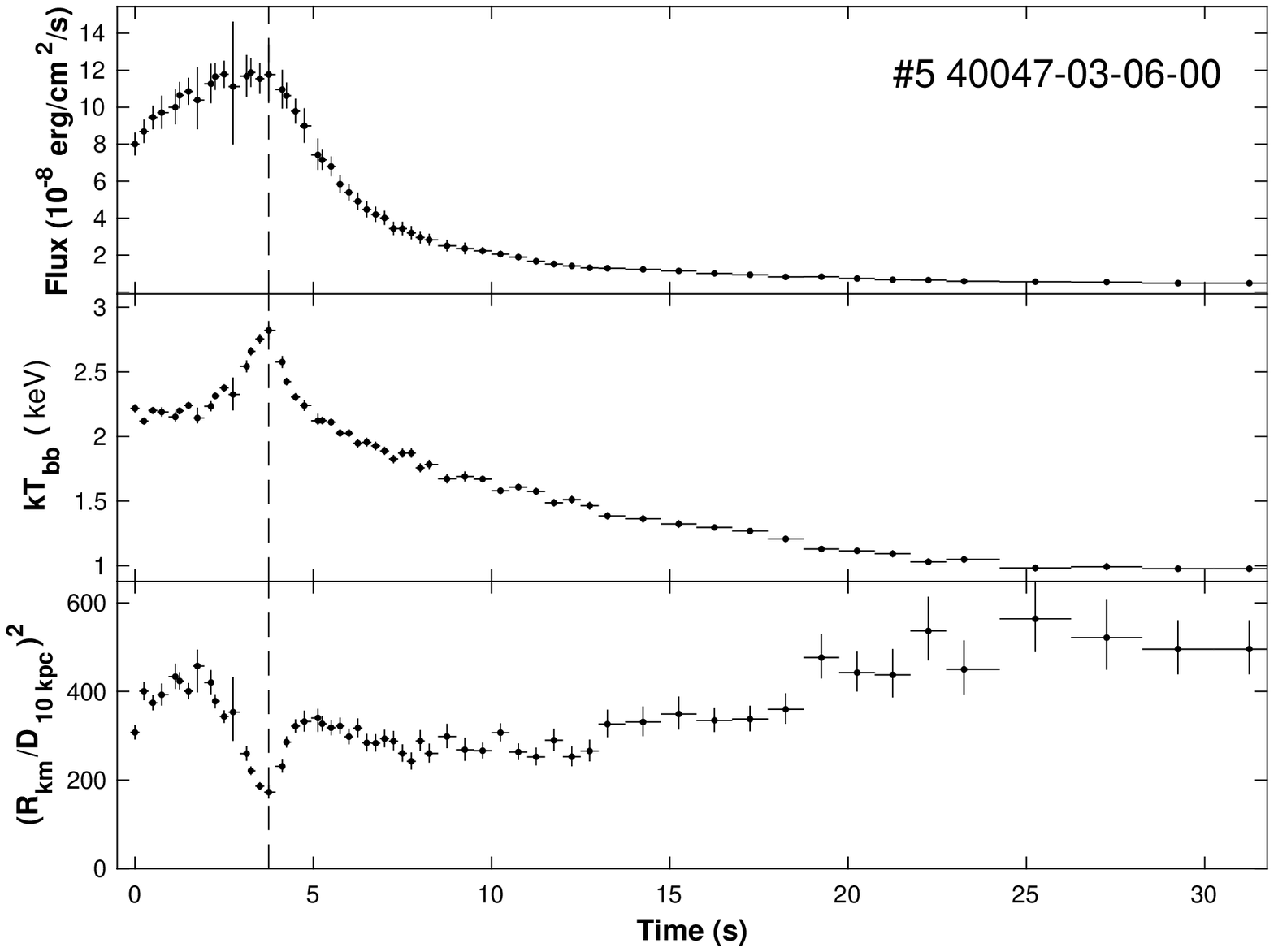}{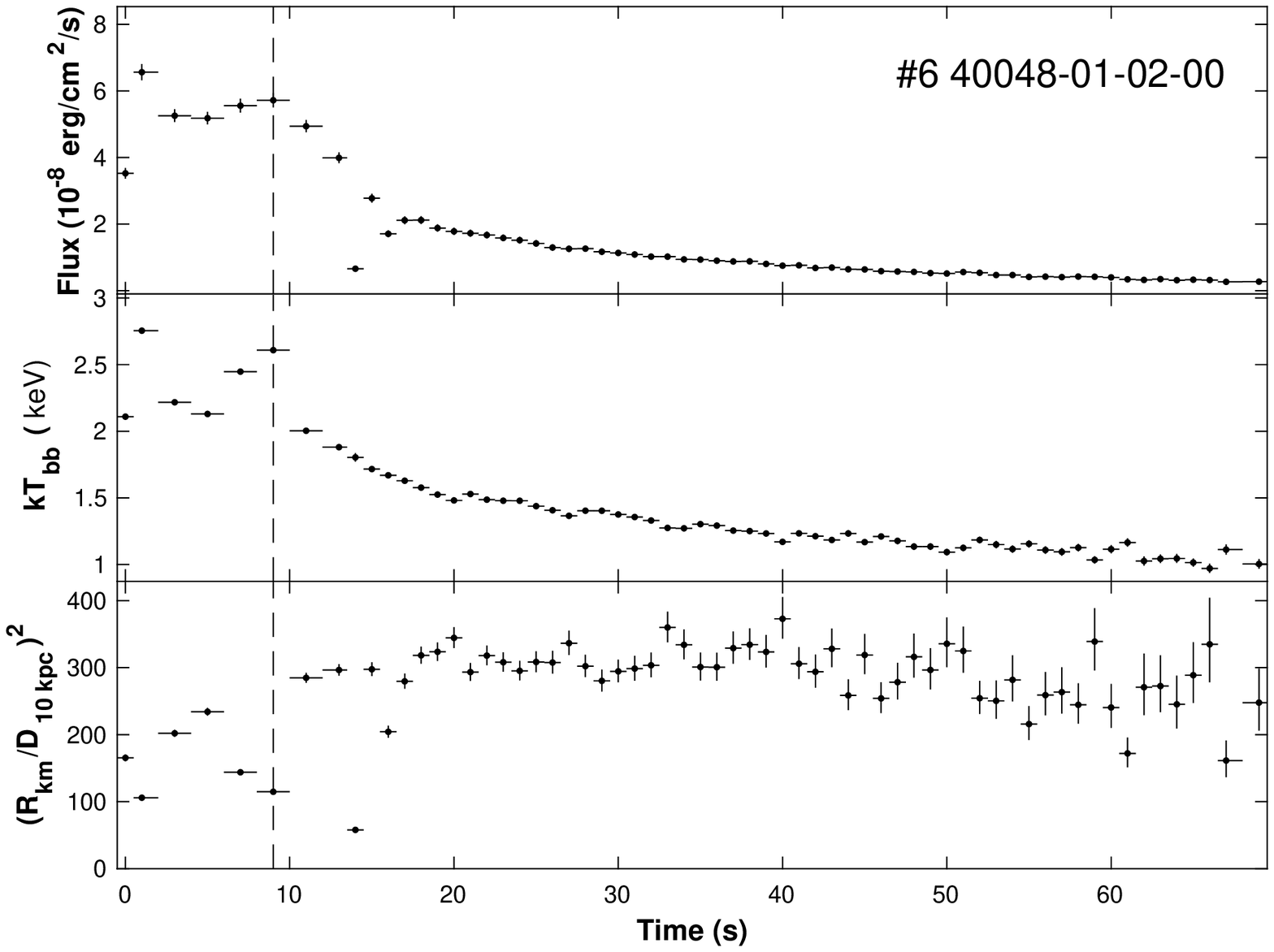}
\plottwo{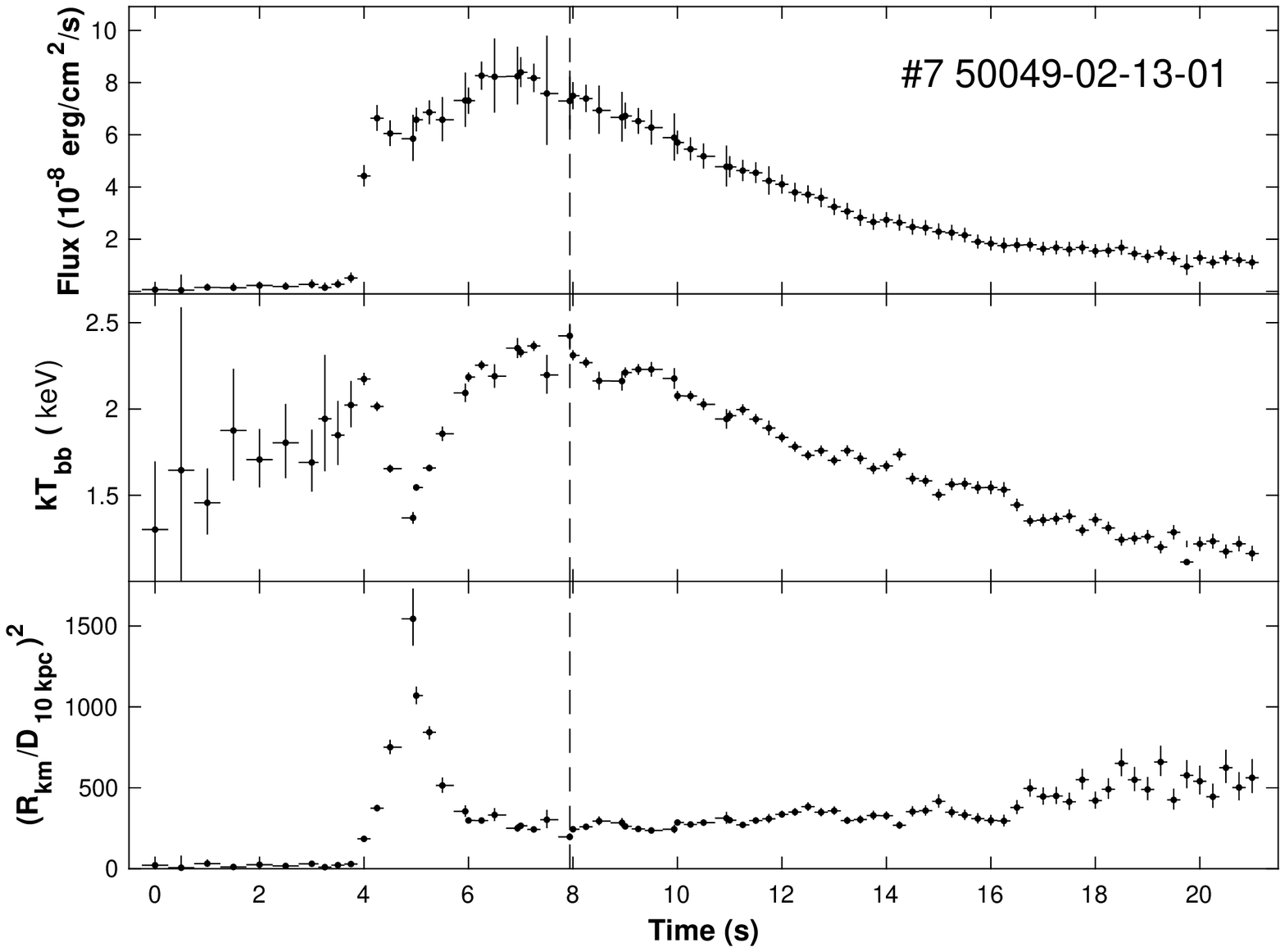}{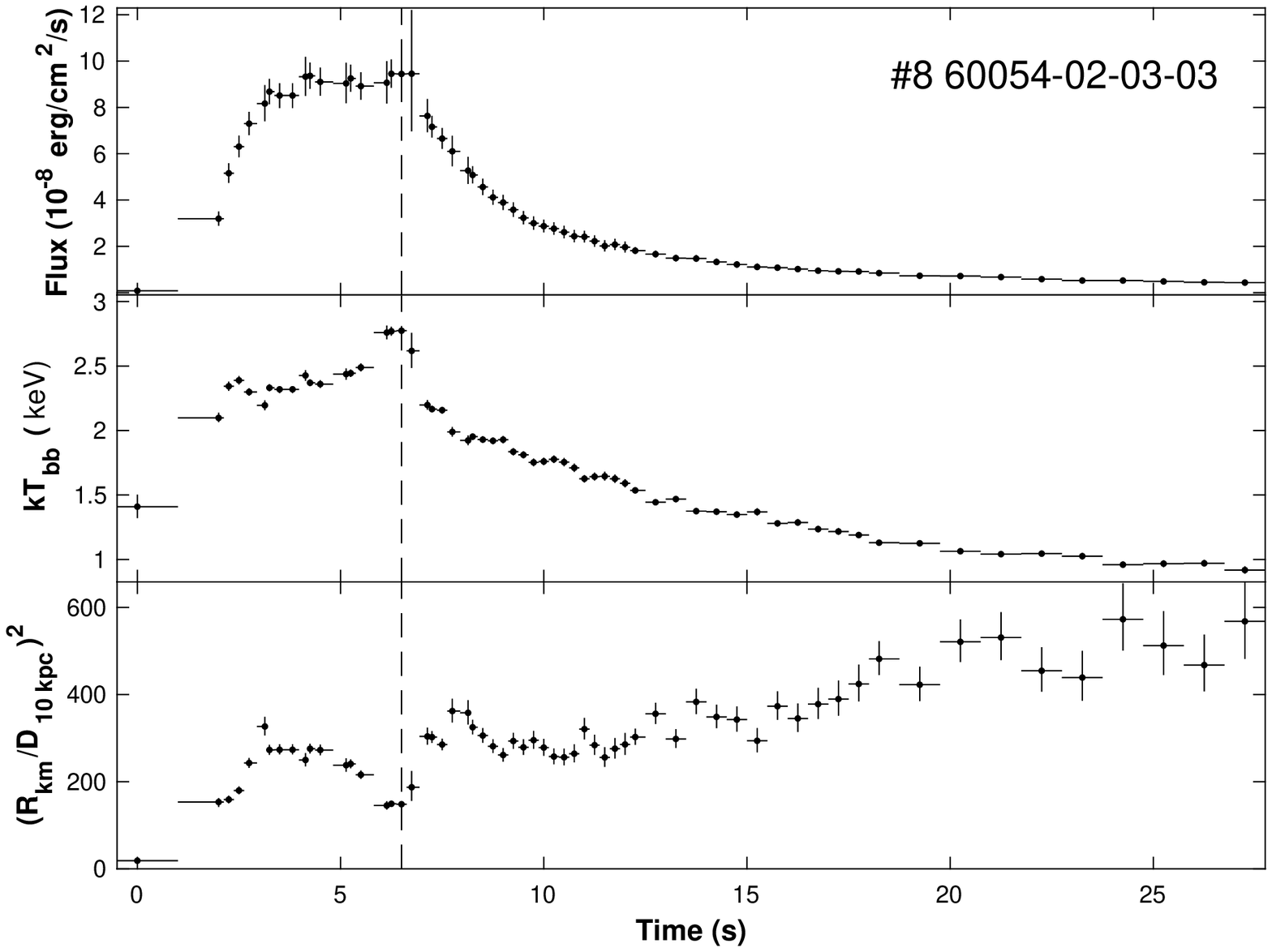}
\caption{Time resolved spectroscopic results from all 14 analyzed PRE bursts. We show the inferred flux from the best fit blackbody model ({\it top panels}); the blackbody temperature ({\it middle panels}); and the blackbody normalization ({\it bottom panels}). The vertical dash line marks the touchdown moment in each plot.  }\label{fig:evol1}
\end{figure*}

\begin{figure*}\centering		
\plottwo{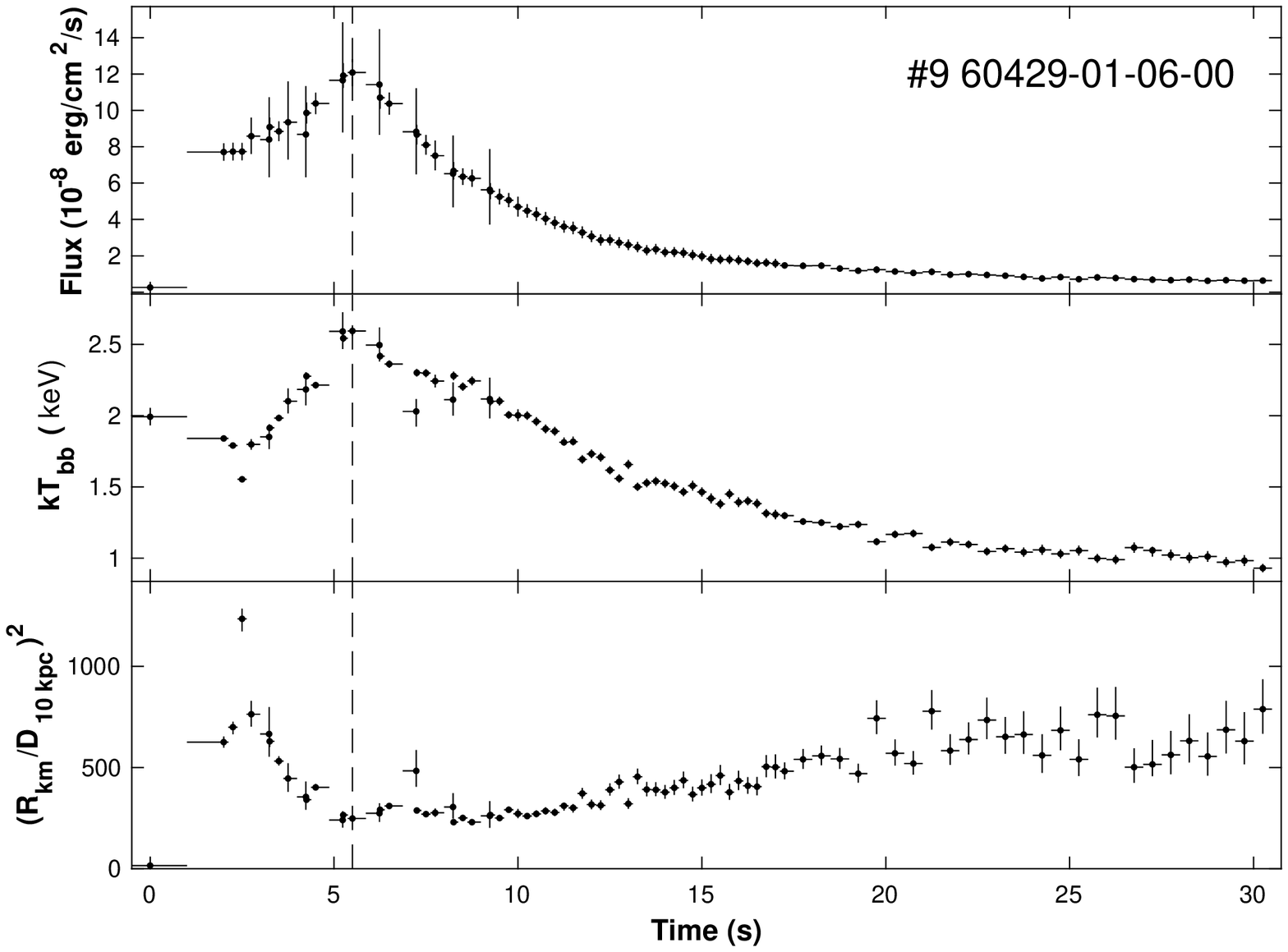}{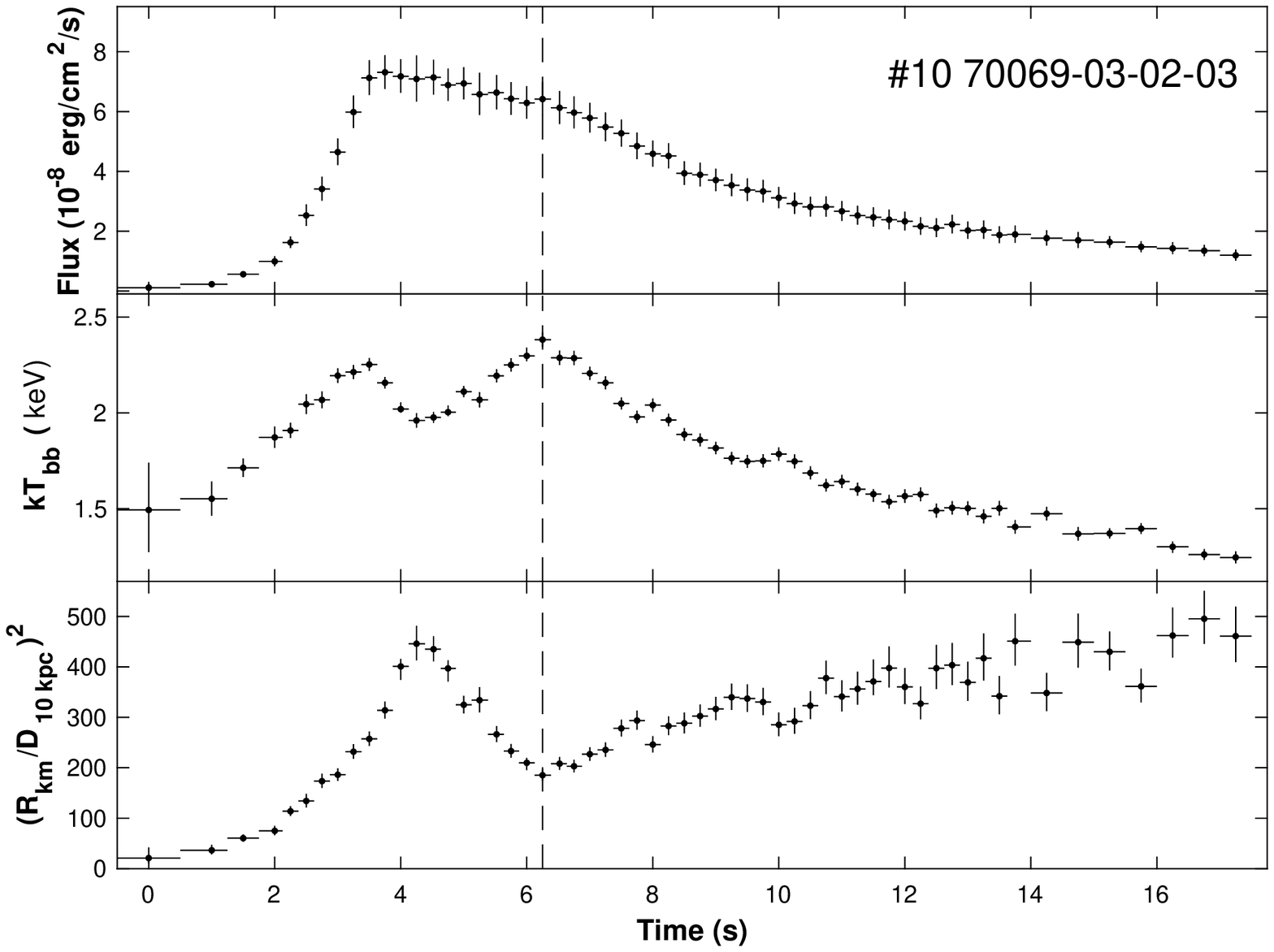}
\plottwo{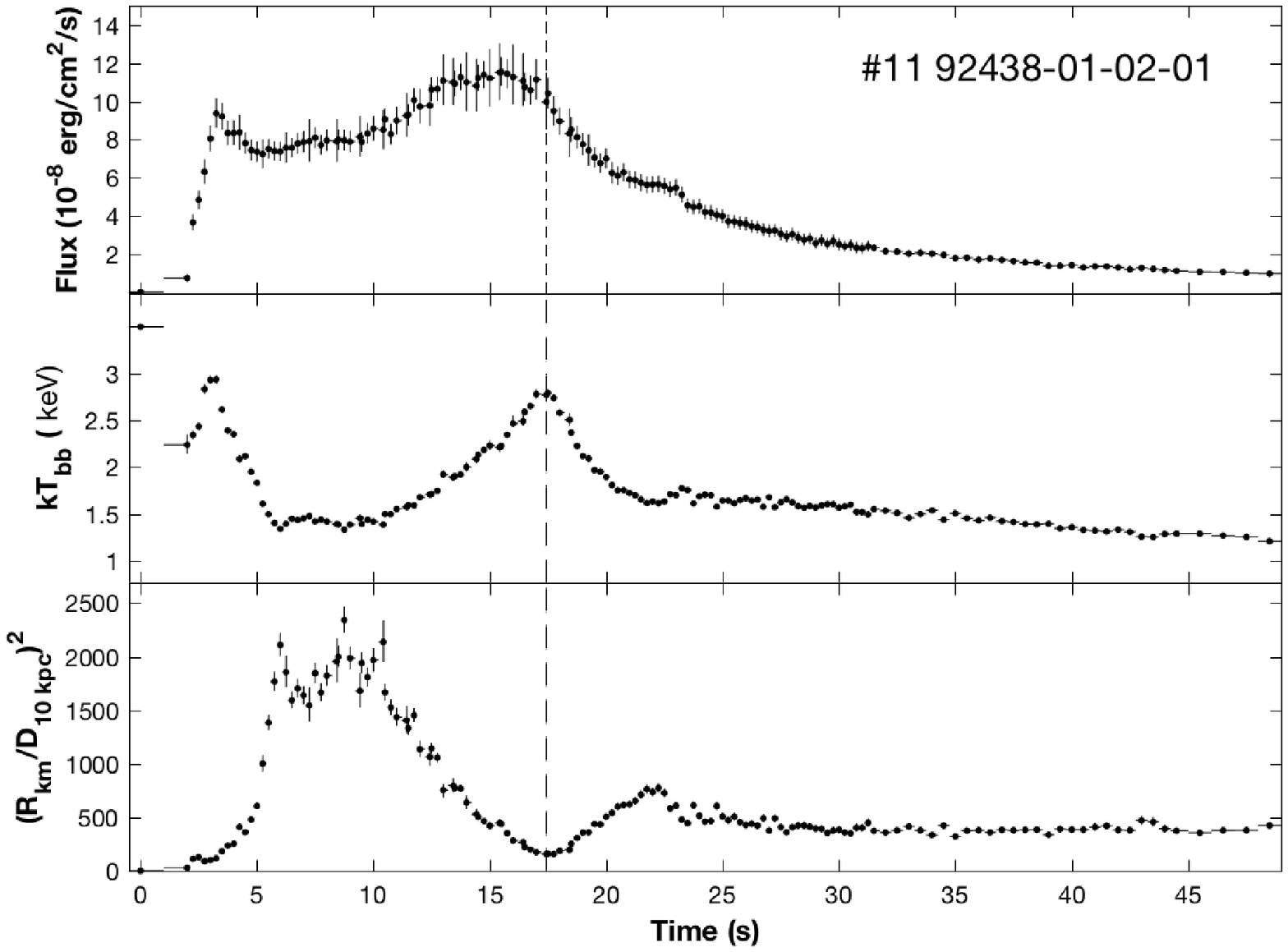}{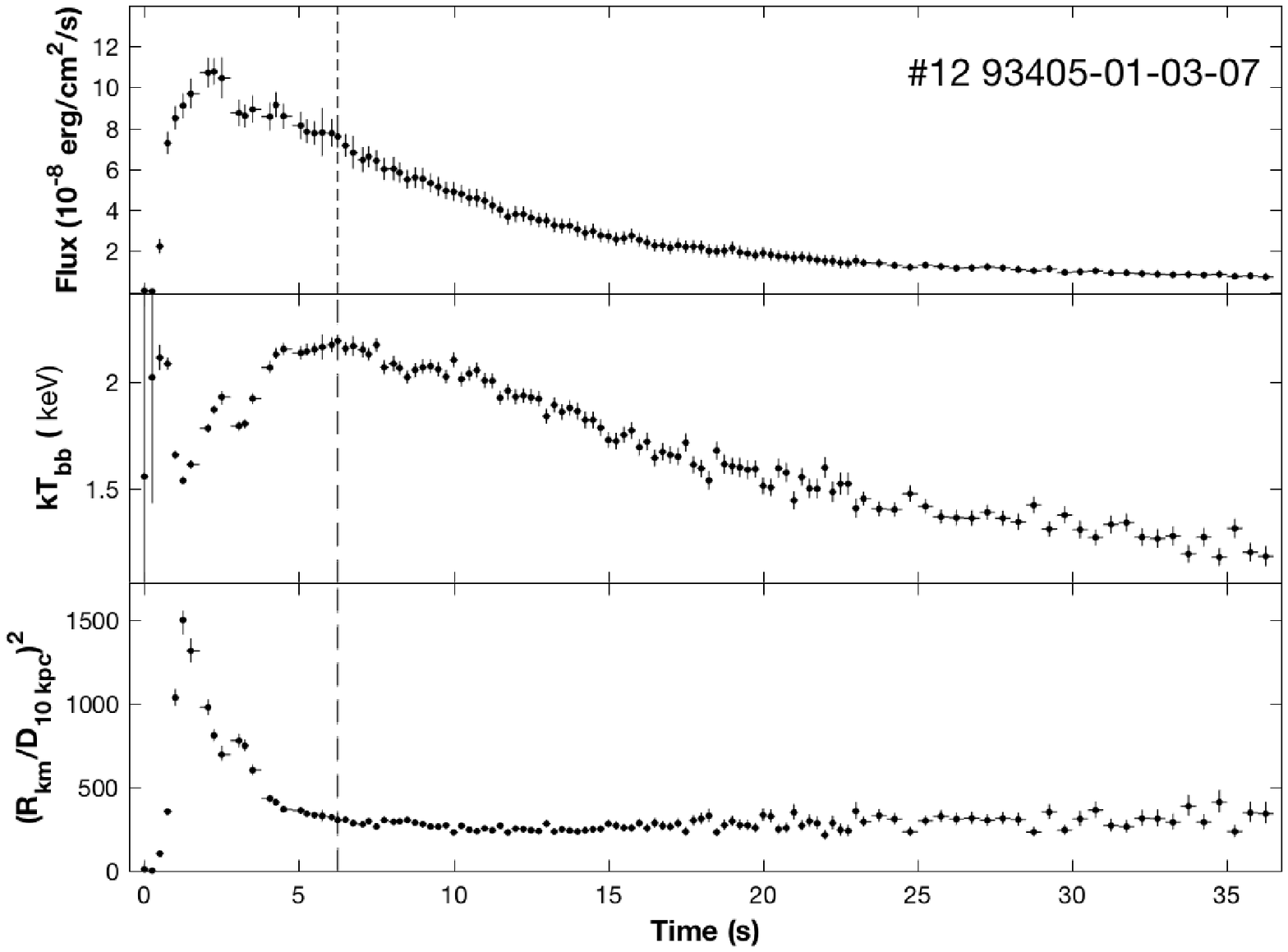}
\plottwo{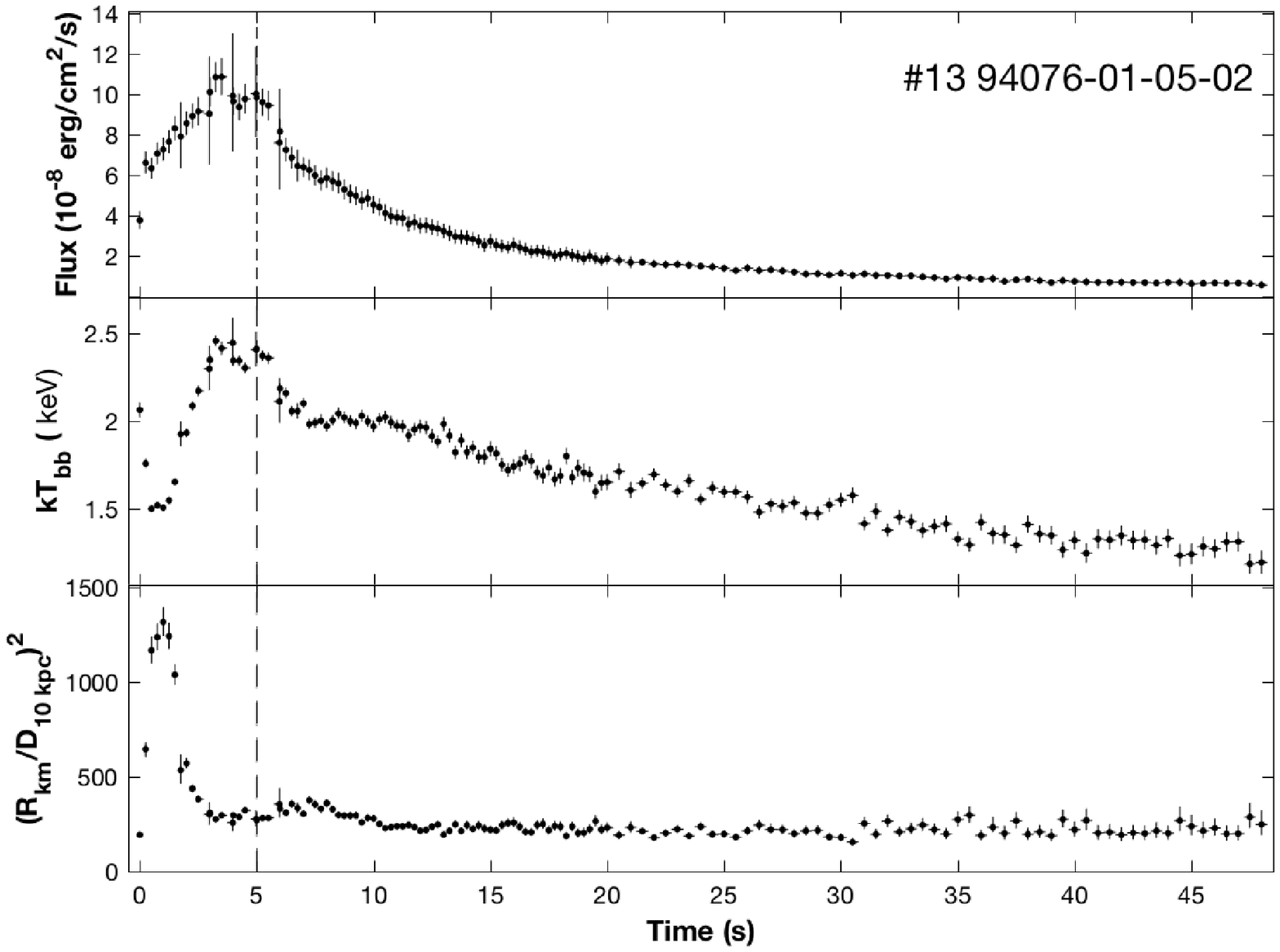}{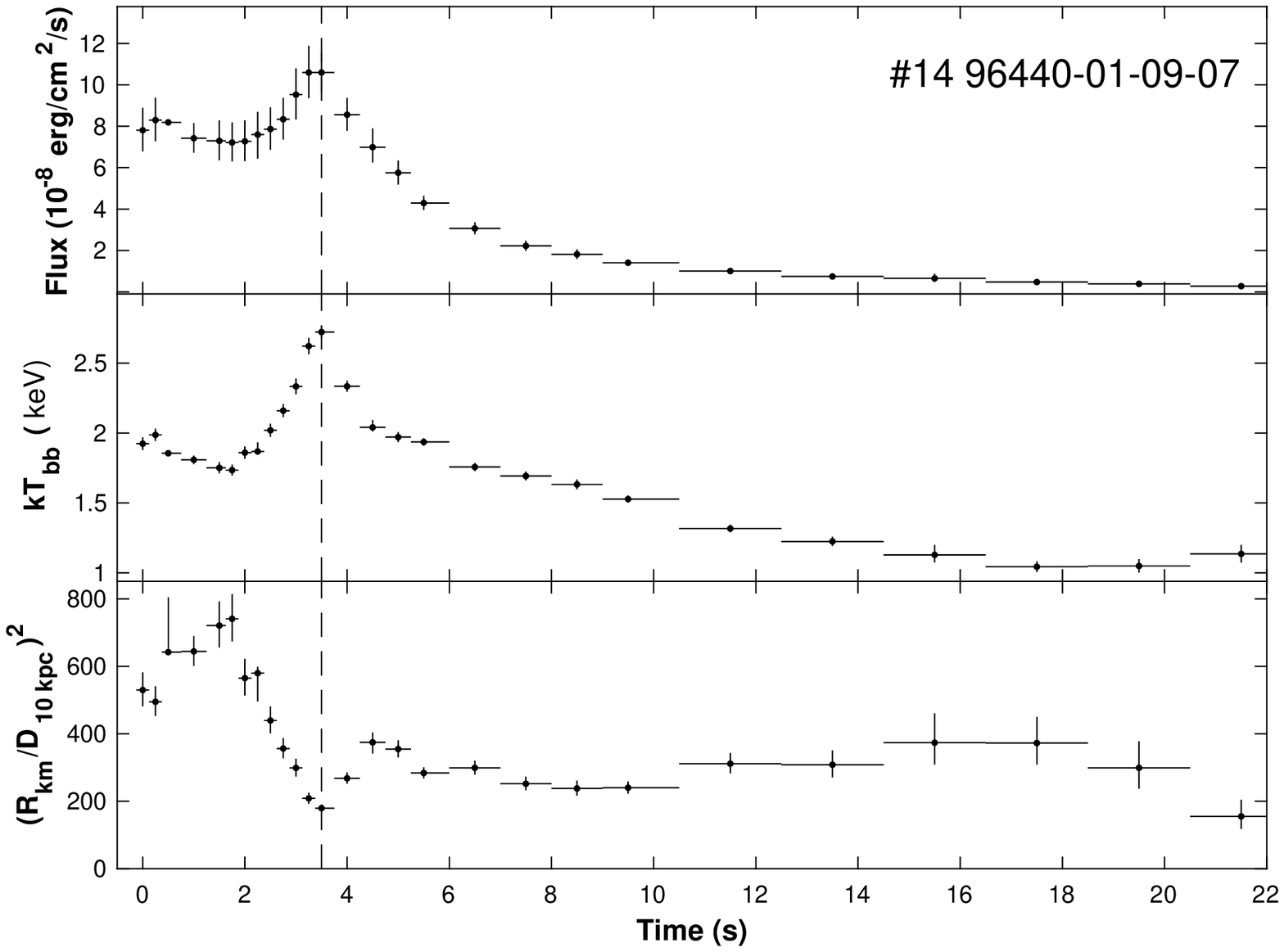}
\caption{Figure~\ref{fig:evol1} continued.}\label{fig:evol2}
\end{figure*}

All 14 PRE type-I X-ray burst spectra were best fitted with an absorbed, {\sc tbabs}, and a blackbody, {\sc bbodyrad}, model. No hard X-ray excesses were recorded in these bursts \citep[see also][]{Galloway08}.  For some of the spectra, a Gaussian component was used to model the broad iron line around 6.4 keV seen in \textit{RXTE}  data. We fixed the Gaussian bandwidth at 0.3 keV to take into account the limited energy resolution of the instrument \citep{Jahoda06}. Given that we were  not able to constrain the hydrogen column density value, $N_{\rm H}$, (as the PCA bandpass starts  above 3 keV) we fixed it to the value of $5.21\times10^{21}~\rm cm^{-2}$ (see Sec. \ref{sec:qspec}). We display the results in Fig.~\ref{fig:evol1}, and~\ref{fig:evol2}.

The bolometric flux is defined as \citep{Galloway08}:
\begin{equation}
F=1.076\times10^{-11}\Bigl(\frac{kT_{\rm bb}}{1~{\rm keV}}\Bigr)^4 K~{\rm  erg\, cm^{-2}\, s^{-1}},
\end{equation}
where the uncertainty was estimated by the general equation of the propagation of the errors.

\begin{figure}
\plotone{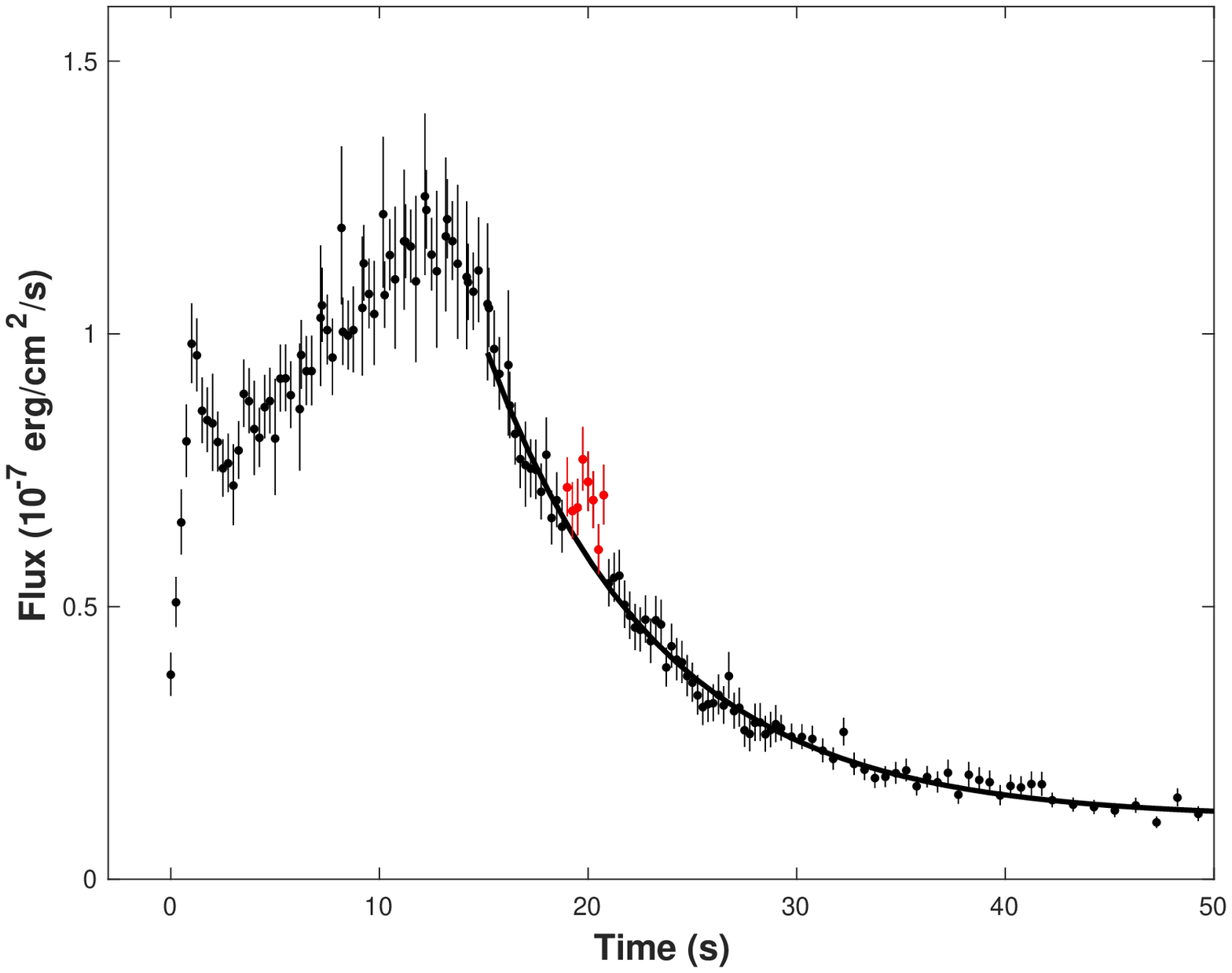}
\caption{PRE burst No. 11 light curve.  The flux excess during the decay phase is highlighted with red data points. The black solid line is the exponential best fit to the decay phase. The fitted function is $Flux=a{\rm e}^{bt}+c$, and the weighted Nelder-Mead method is applied, where $a=5.34\times10^{-7},~ b=-0.12, ~ c=1.4\times10^{-8}$.  }\label{fig:decay}
\end{figure}

The burst No. 6 (Obs ID: 40048-01-02-00) is a peculiar event, as it has been identified as a PRE burst by \citet{Galloway08} but not by \citet{Chen13} and by \citet{Kajava14}. In addition, as its time-solved spectra frequently show null values, we excluded it for the  \object{Aql X--1}  mass and radius determination procedure. For completeness, we just show this peculiar burst with a time bin of 2 s to avoid null bin values in Fig.~\ref{fig:evol1}.  Burst No. 11 shows a flux excess during its cooling tail, i.e., occurring $t=20~{\rm s}$ from the onset of the burst. This flux excess is highlighted in red in Fig.~\ref{fig:decay} and excluded for the following analysis in this work. 

\begin{deluxetable*}{lccccccc}
\tabletypesize{\scriptsize}
\tablecaption{Log of the Aql X--1 {\em RXTE} PRE bursts observations analyzed in this paper and spectral properties of the bursts.
\label{table:burst}}
\tablewidth{0pt}
\tablecolumns{5}
\tablehead{\colhead{Obs\_ID} &\colhead{Burst No.\tablenotemark{1}}      & \colhead{Touchdown flux } & \colhead{Peak flux} &\colhead{$K$ ($0.1-0.7F_{\rm TD}$)\tablenotemark{2}} & \colhead{$kT_{\rm TD}$\tablenotemark{3}}    &\colhead{PCU\tablenotemark{4}}  &\\
			\colhead{}     &  & \colhead{$10^{-7}~{\rm erg\, cm^{-2}\,s^{-1}}$}  &  \colhead{$10^{-7}~{\rm  erg\, cm^{-2}\,s^{-1}}$}     & \colhead{$({\rm km /10~kpc})^2$}    &\colhead{(keV)} &   &    }
		\startdata
		20092-01-05-00  & {\#}1 (5) &  $1.20\pm0.08$ & $1.20\pm0.08$   & $  278\pm   30$ & $2.52\pm0.02$ &All& \\	
		20092-01-05-030 & {\#}2 (6)  &  $ 0.62\pm0.09$ & $0.74\pm0.15$  & $280  \pm 41$ &$2.28\pm0.06$ & All& \\	
		20098-03-08-00  & {\#}3 (4)  &  $1.20\pm0.06$ &  $1.20\pm0.06$   & $321  \pm 101$&$2.66\pm 0.04$ &All& \\	
		40047-03-02-00  & {\#}4 (10)  & $1.29\pm 0.07$ & $1.29\pm0.07$  & $272 \pm  26$ &$ 2.52\pm0.03$ &0,2,3,4&\\	
		40047-03-06-00  & {\#}5 (11)  & $1.18\pm0.16$ & $1.19\pm 0.78$   & $290  \pm 29$&$ 2.82\pm0.07$ &0,2,4& \\	
		40048-01-02-00  & {\#}6 (13)  &$0.57\pm0.02$ & $0.66\pm0.02$   & $290   \pm49$&$2.61\pm 0.02$ &0,1,2,3& \\	
		50049-02-13-01  &  {\#}7 (19)  &$0.73\pm0.11$ &$0.84\pm0.06$    & $376 \pm 113$&$2.42\pm 0.07$  &0,2,3,4&\\	
		60054-02-03-03  &  {\#}8 (25) &$0.94\pm0.06$  &$0.95\pm0.25$    & $304 \pm  38$&$2.16\pm 0.03$ &0,1,2,3& \\	
		60429-01-06-00  &  {\#}9 (28)  &$1.21\pm0.08$ &$1.21\pm0.08$    & $391  \pm124$&$ 2.59\pm 0.03$ &0,2,3,4& \\	
		70069-03-02-03  &  {\#}10 (29)  &$0.64\pm0.06$ &$0.73\pm0.06$   & $339  \pm 77$&$ 2.38\pm 0.04$ &0,2,3& \\	
		92438-01-02-01  &  {\#}11   &$1.03\pm0.13$& $1.18\pm0.07$   & $584 \pm 158$&$2.77\pm 0.07$ &0,2,4& \\ 		
		93405-01-03-07  &  {\#}12   &$0.76\pm0.06$ &$1.08\pm0.06$   & $277 \pm  33$&$2.20\pm 0.03$ &0,1,2& \\		
		94076-01-05-02  &  {\#}13   &$1.00\pm0.21$ &$1.08\pm0.09$   & $242\pm   45$ &$ 2.19\pm 0.03$ &1,2,4&\\		
		96440-01-09-07  &  {\#}14 & $1.06\pm0.09$ & $1.06\pm0.09$ & $291 \pm  55$  &$2.72\pm 0.05$ &2&
		\enddata
		\tablenotetext{1}{Burst No. marked in this work. In parentheses, it is the Burst No. marked in  \citet{Galloway08}. The PRE bursts in last four observations were identified lately. }
	        \tablenotetext{2}{The mean value and standard deviation of apparent angular size in the range $0.1-0.7F_{\rm TD}$ during the cooling tail. Here, $F_{\rm TD}=1.06\times10^{-7}~{\rm erg\, cm^{-2}\, s^{-1}}$.}
	        		\tablenotetext{3}{The blackbody temperature at the touchdown moment.}

		\tablenotetext{4}{The active Proportional Counter Units (PCUs) during the burst epoch.}

	\end{deluxetable*}
 
\section{Results}
\label{sec:results}
\begin{figure*}\centering
\includegraphics[angle=270,scale=0.35]{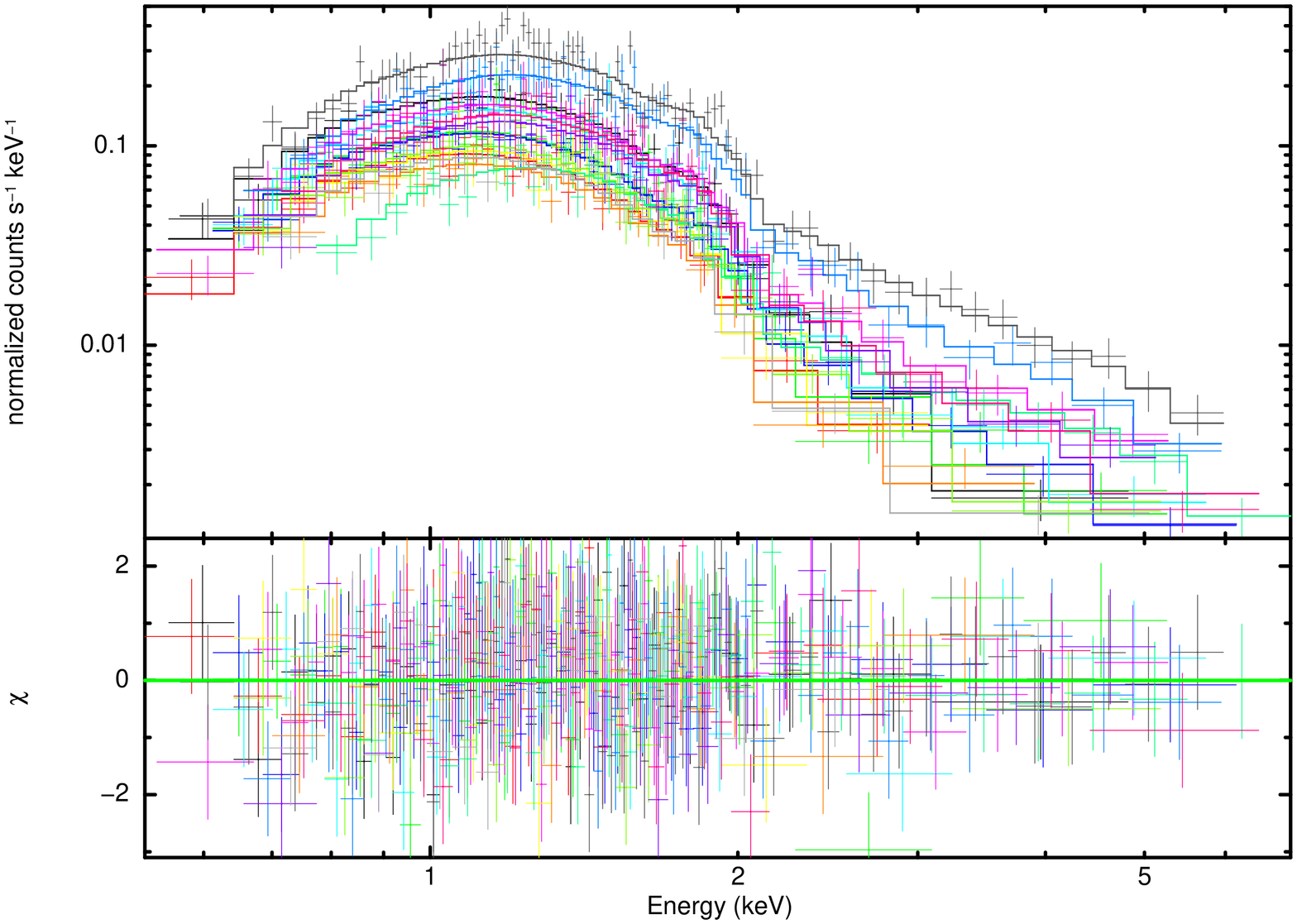}	
\includegraphics[angle=270,scale=0.35]{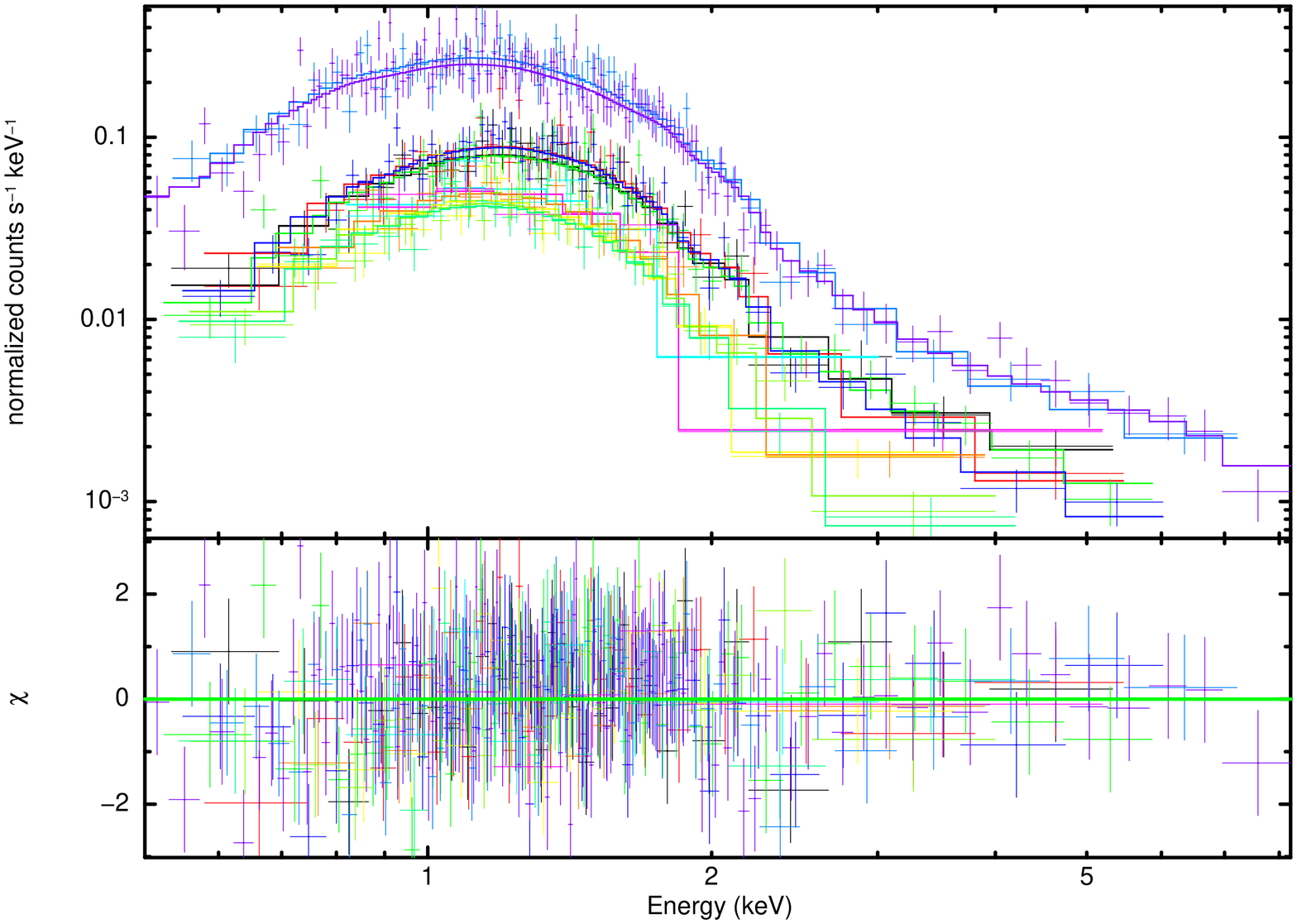}
\caption{Left panel Aql X--1 spectra observed by \textit{Chandra}. Right panel we show the \textit{XMM-Newton} spectra. We plot the absorbed spectrum of Aql X--1 fit with an absorbed, {\sc tbabs}, NS thermal atmosphere, {\sc nsatmos}, and power-law model.  The $N_{\rm H}$ value, the source distance, the NS mass and radius are fixed at $5.21\times10^{21}~{\rm cm^{-2}}$, 5 kpc, $1.4M_{\odot}$ and $10 ~{\rm km}$, respectively. The lower panel shows the residual between the data and the model.}
\label{fig:quiescent_spectra}
\end{figure*}
	
\subsection{The quiescent spectra}
\label{sec:qspec}	
The spectral analysis was carried out using  {\sc xspec} version 12.8.2 \citep{Arnaud96}. We studied in detail the quiescent X-ray spectra of  Aql X--1 in the energy range $0.5-10~{\rm keV}$, using \textit{Chandra}/ACIS-S and  \textit{XMM-Newton}/PN/MOS data. We grouped all \textit{Chandra} and \textit{XMM-Newton} spectra in order to have at least 20 photons in each energy bin. We summarize in Table \ref{table:quiescent} all the analyzed spectra. To account for the soft thermal and hard emission components, we fitted all the spectra with an absorbed NS atmosphere, {\sc nsatmos}, adding also a power-law component. The free parameters are the hydrogen column density, $N_{\rm H}$, NS atmosphere temperature, the distance to the source, the NS mass and radius, the power-law index, $\Gamma$ and the power-law normalization. We assume throughout the spectral fits that during the quiescent state the whole NS surface is radiating. Therefore, the {\sc nsatmos} emission fraction parameter was set to unity. We used the {\sc xspec} absorption model, {\sc tbabs}, with the {\sc wilm} abundances \citep{Wilms00}. All uncertainties in the spectral parameters are given at  $1\sigma$ c.l. for a single parameter.  

To compare the consistence due to the absolute flux calibration between \textit{XMM-Newton}/PN/MOS and \textit{Chandra}/ACIS-S spectra \citep[see e.g.,][]{Guver15}, we first fitted separately all the \textit{XMM-Newton}/PN/MOS and \textit{Chandra}/ACIS-S spectra. For the {\sc nsatmos} model we let free to vary the atmosphere temperature and we fixed the NS mass and radius to the canonical values ($M = 1.4 M_{\odot}$, $R=10$~km). The source distance was first set to the lower limit of 4 kpc and afterwards to the upper limit of 6.25 kpc. For the lower and upper source distance values, the best fits provided an absorption column density of $N_{\rm H}=(6.1-5.1)\pm0.1 \times 10^{21} \,{\rm cm}^{-2}$, a power-law photon index $\Gamma=(1.2-0.7)\pm0.2$ with a $\chi^{2}_{\rm red}{\rm /d.o.f.} = 0.9/48$1 and 0.9/591 for \textit{XMM-Newton}/PN/MOS and \textit{Chandra}/ACIS-S  spectra, respectively. For the joint \textit{XMM-Newton} data we include a normalization constant in the fit to take into account the uncertainties in the cross-calibrations of the instruments and the source variability (the data are not covering strictly the same time interval). In this fit, the normalization of the MOS data was fixed to unity as a reference, while the normalization of the PN data was found to be $1.1 \pm 0.2$. Similar values were found also for the combined \textit{XMM-Newton}/PN/MOS and \textit{Chandra}/ACIS-S spectra fit with a $\chi^{2}_{\rm red}{\rm /d.o.f.} = 0.9/1073$. In this case the normalization constant  was fixed to unity for the \textit{Chandra} data,  while the normalization constants of the PN and MOS data was found to be $0.9 \pm 0.1$, for  both instruments. The quiescent spectra from \textit{Chandra} and \textit{XMM-Newton} and their best fitting models are displayed in Figure~\ref{fig:quiescent_spectra}.

\begin{figure}
\epsscale{1}
\plotone{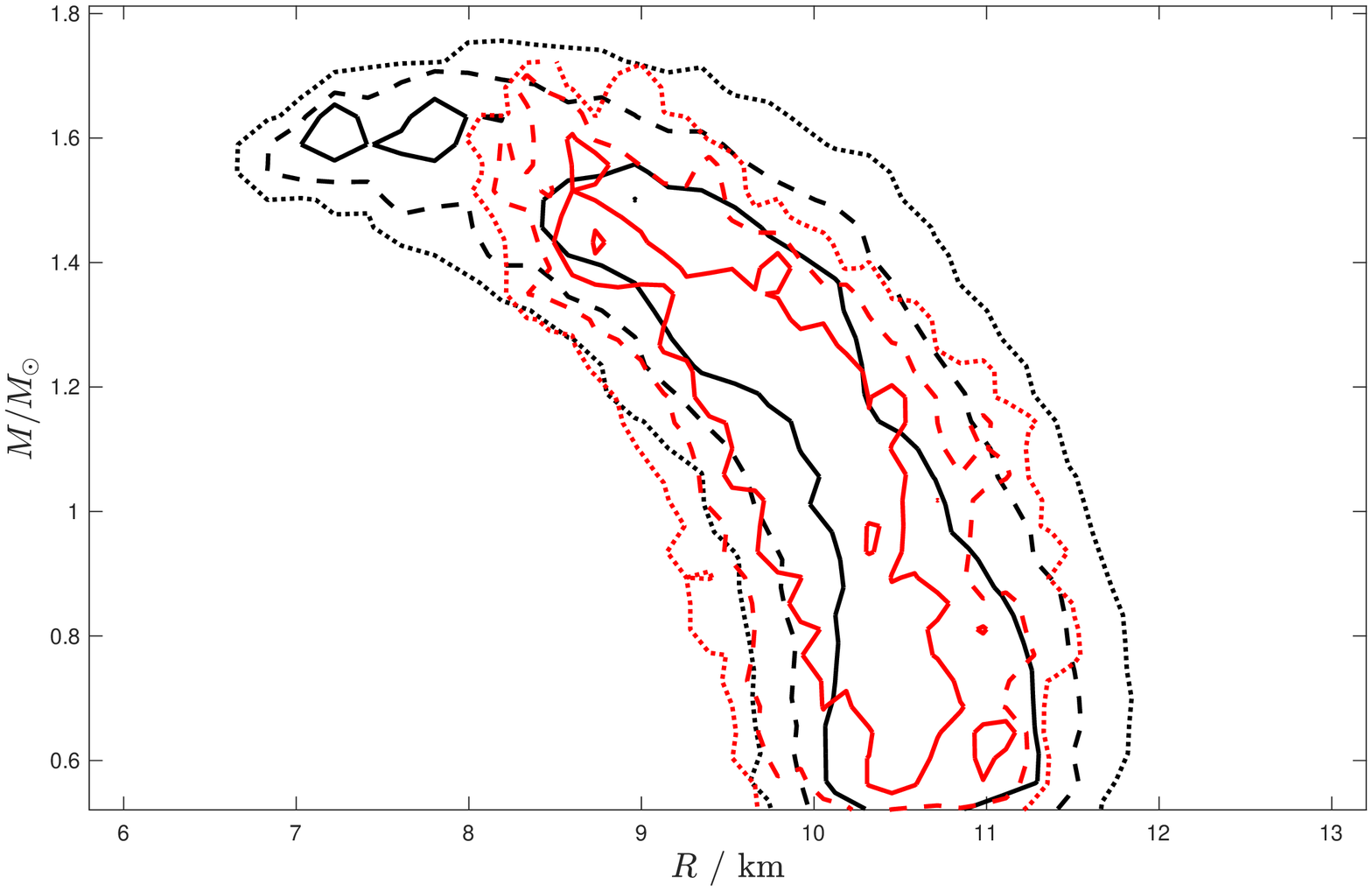}
\caption{We show the consistency of the $M-R$-relation between the  \textit{Chandra} and \textit{XMM-Newton observations}. The 1, 2, 3$\sigma$ confidence regions of the mass and radius in Aql X--1 from \textit{Chandra} (red contours) and \textit{XMM-Newton} (black contours). The distance is $5\pm0.25$ kpc and the $N_{\rm H}$ is fixed at $5.21\times10^{21}~{\rm cm^{-2}}$. For more details, please see Section~\ref{sec:mass-radius}.} \label{fig:chandra_newton}
\end{figure}

Although the fits were all formally acceptable, the NS mass and radius have to be free parameters. We thus fixed throughout this work the model independent reported hydrogen column density at  $5.21\times10^{21}~{\rm cm^{-2}}$ \citep[][see Sec. \ref{sec:source}]{Pinto13}. We then fitted again the \textit{Chandra} and \textit{XMM-Newton} spectra separately to investigate the consistence of the mass and radius constraints. The distance was set at $5\pm0.25~ {\rm kpc}$.  The NS atmosphere temperature  is a free parameter for all spectra, while the power-law index were tied together and free to change for \textit{Chandra} and \textit{XMM-Newton} spectra.  
To consider now a mass and radius skewed distribution we apply the Goodman-Weare algorithm of Monte Carlo Markov Chain \citep[MCMC; e.g.,][]{Guillot13,Goodman10}. This simulation procedure is implemented as {\sc chain} in the {\sc xspec} package. For each of the 200  chains, the length was $2\times10^6$. The first 20\% of the simulated data were burned. The mass and radius contours are displayed in Figure~\ref{fig:chandra_newton}. 

The mass and radius of the NS constrained from \textit{Chandra} and \textit{XMM-Newton} are consistent with each other. However, 
\textit{Chandra} spectra provide tighter constraints on the NS mass and radius for \object{Aql X--1} than \textit{XMM-Newton}. The most likely explanation is that \textit{Chandra} observed \object{Aql X--1} more frequently than \textit{XMM-Newton} as listed in Table~\ref{table:quiescent}. The total number of photons collected by \textit{Chandra} is about 1.34 times larger than \textit{XMM-Newton},  so the signal-to-noise ratio of the \textit{Chandra} spectra is higher than that of the \textit{XMM-Newton} spectra shown in Figure~\ref{fig:quiescent_spectra}.  We note that the nicely overlapping mass-radius contours are not obtained if we set the hydrogen column density at $N_{\rm H}=6.1 \times 10^{21} \,{\rm cm}^{-2}$. We thus preferred to use the model independent determined $N_{\rm H}$ value of $5.21\times10^{21}~{\rm cm^{-2}}$. 

We directly combined the \textit{Chandra} and \textit{XMM-Newton} observations together and derived the mass-radius relation in Sec \ref{sec:mass-radius}. 

\subsection{PRE bursts in Aql X--1}
\label{sec:pre}

\subsubsection{Theory-driven approach}

For each PRE burst we fitted the $F-K^{-1/4}$ relation obtained from the observations with the theoretical model No. 8 from \citet{Suleimanov12}, where $\log g=14.3$, and $X=0.7343,~Y=0.2586$, $Z=4.02\times 10^{-3}$ are the abundances of hydrogen, helium, and metals, respectively.  The purpose of the selected abundance is that the burst cooling atmosphere probably has similar composition as its main sequence companion star. We consider first that the model predicted PRE burst Eddington flux, $F_{\rm Edd}$, is free to vary between $0.8-1.2$ times the touchdown flux, and then that only cooling tail fluxes larger than $F_{\rm TD}/e$ are considered \citep{Poutanen14}. We obtain the best-fit results with the regression method by minimizing the function \citep[see e.g.,][]{Deming11, Suleimanov17}: 
\begin{equation}
\chi^2  =   \sum_{i=1}^{N_{\rm obs}}\left[ \frac {(A f_c - K_i^{-1/4})^2} {(\sigma_{K_i^{-1/4}})^2}+ \frac{( F_{\rm Edd} \ell- F_{\rm i})^2}{\sigma_{F_{\rm i}}^2}\right].
\end{equation}
The $K_i$ and $F_i$ are the $i$th data points in the cooling tail,  where $\sigma_{K_i^{-1/4}}$ and $\sigma_{F_{\rm i}}$ are the corresponding errors. The theoretical relation $\ell-f_c$ is adapted from \citet{Suleimanov12}. The errors of the data are taken into account in both directions. The term in the bracket is the square of the normalized distance from the $i$th data to the model curve $\ell-f_c$. 

\begin{figure}
\plotone{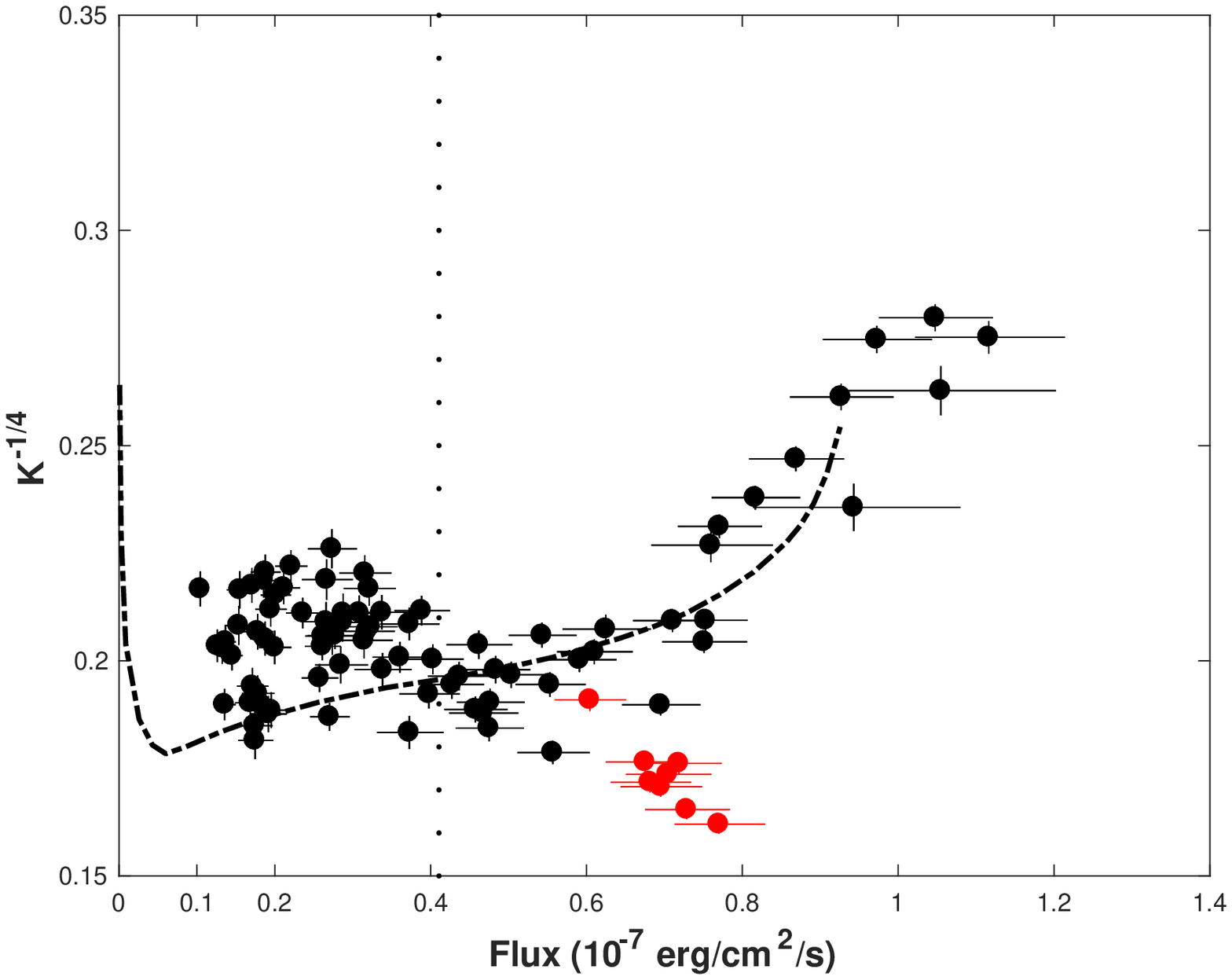}
\plotone{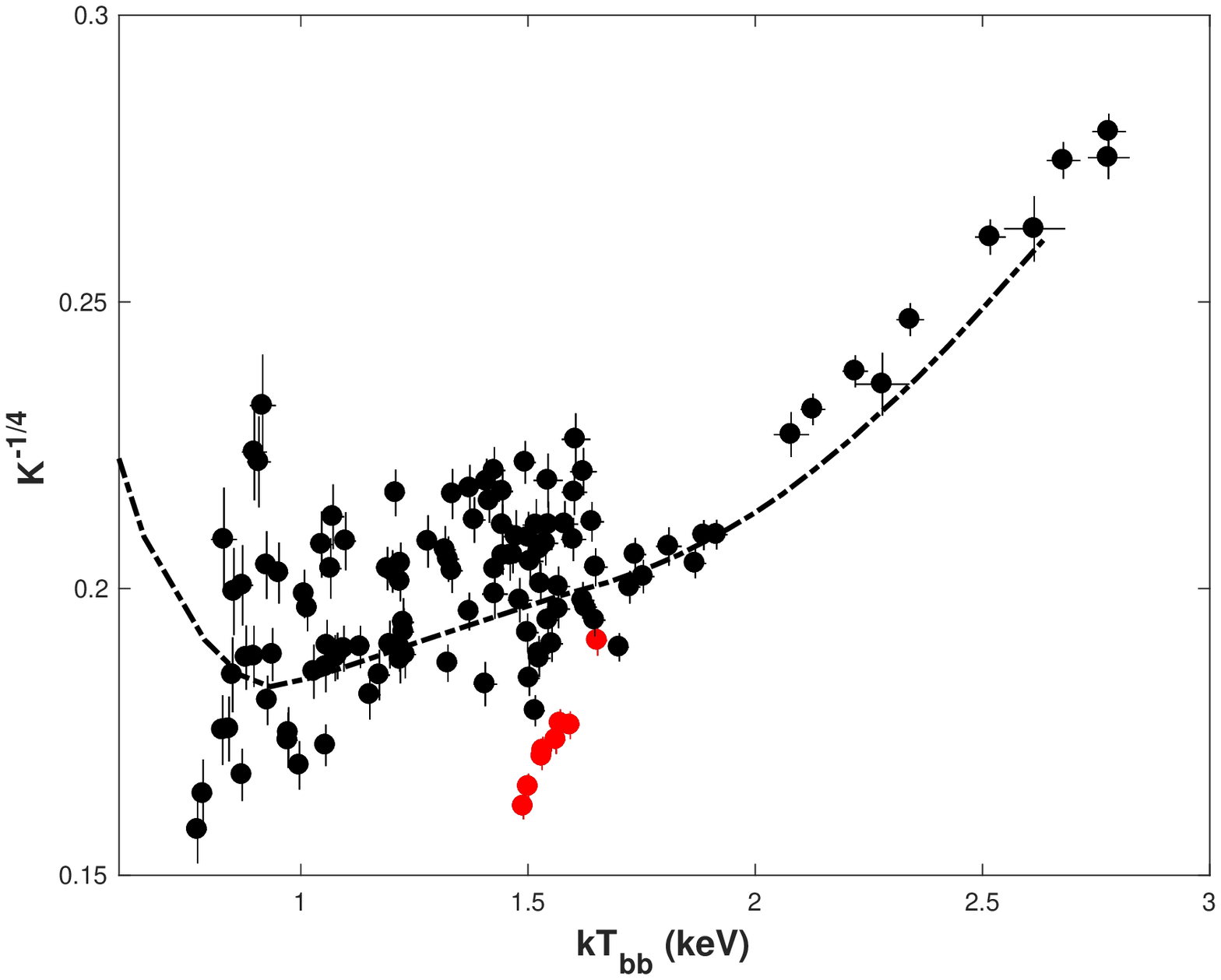}
\caption{The $F-K^{-1/4}$ (top panel) and  $kT_{\rm bb}-K^{-1/4}$ tracks of the hard state PRE burst  (No. 11) in Aql X--1. The red dots are the flux excess as shown in Fig.~\ref{fig:decay}, which are excluded in our fits. }\label{fig:cooling_hard}
\end{figure}

\begin{figure}\centering
\plottwo{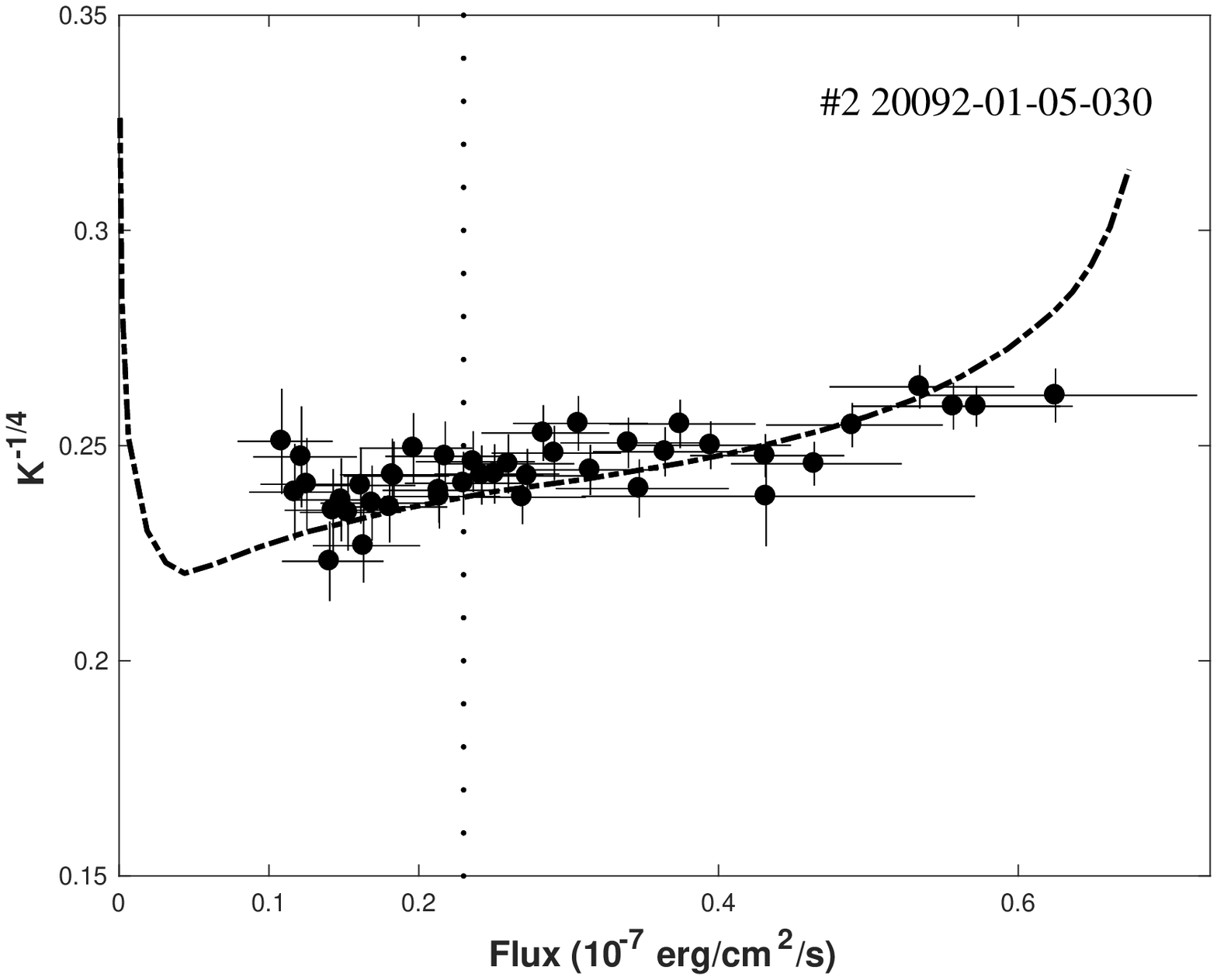}{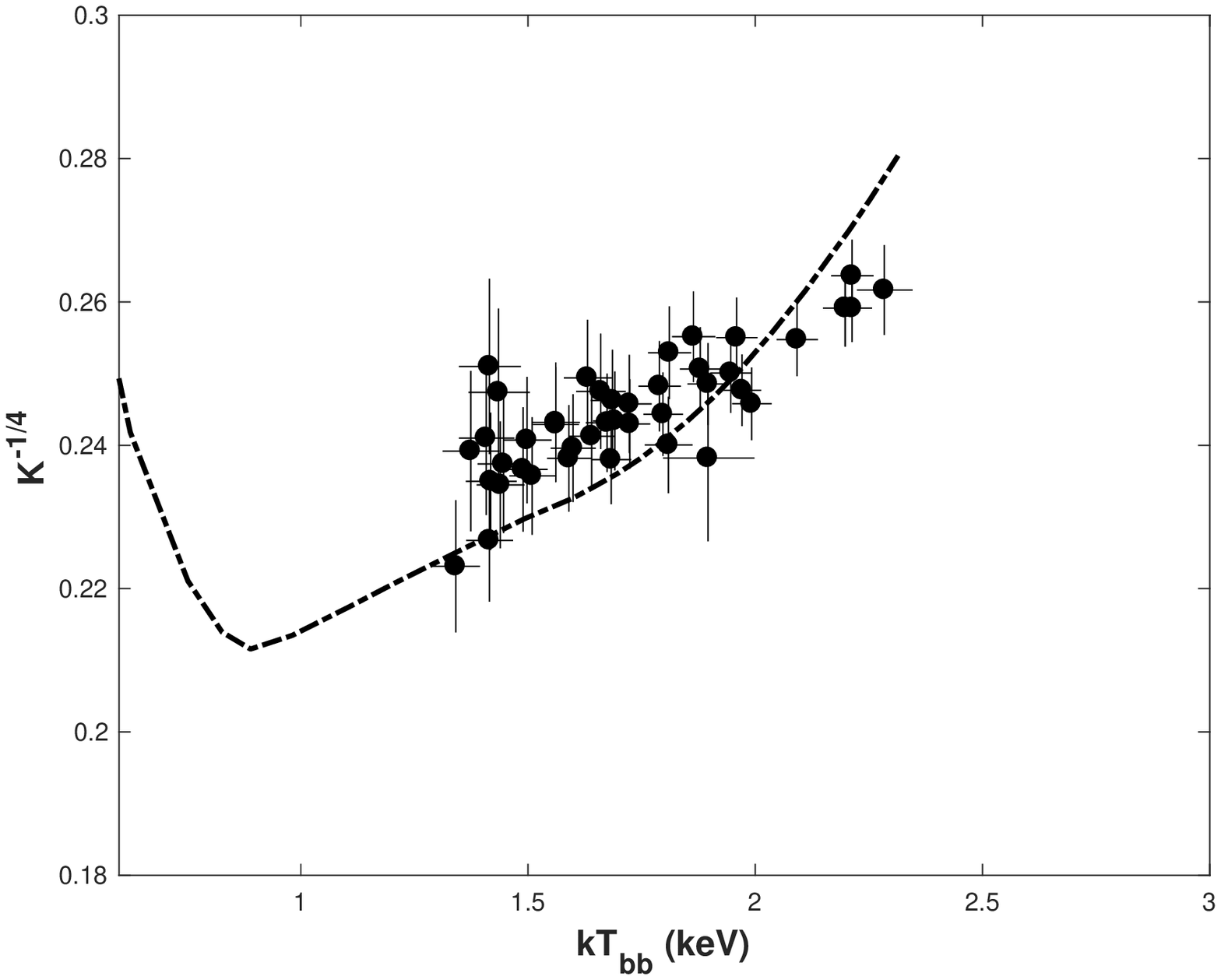}
\plottwo{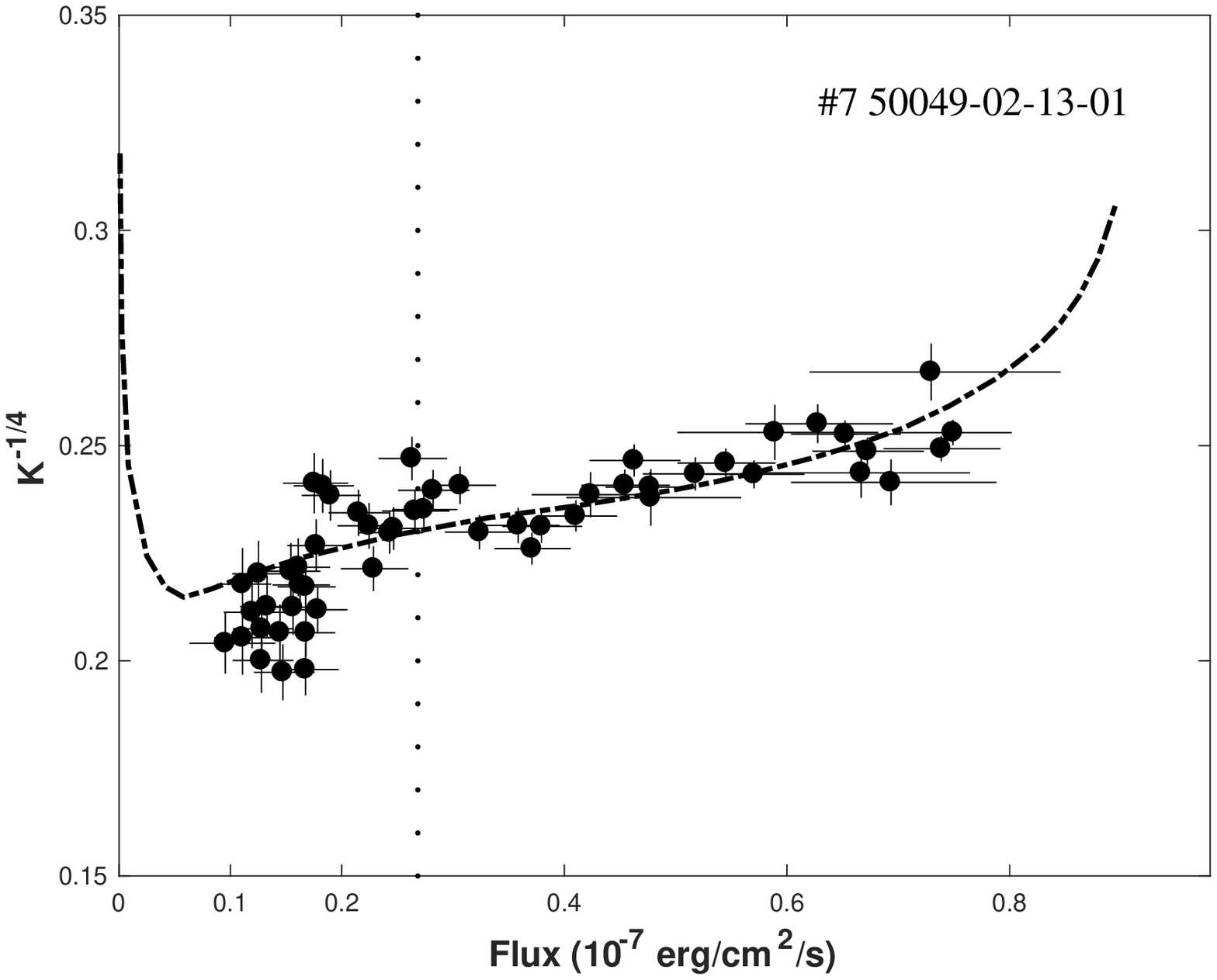}{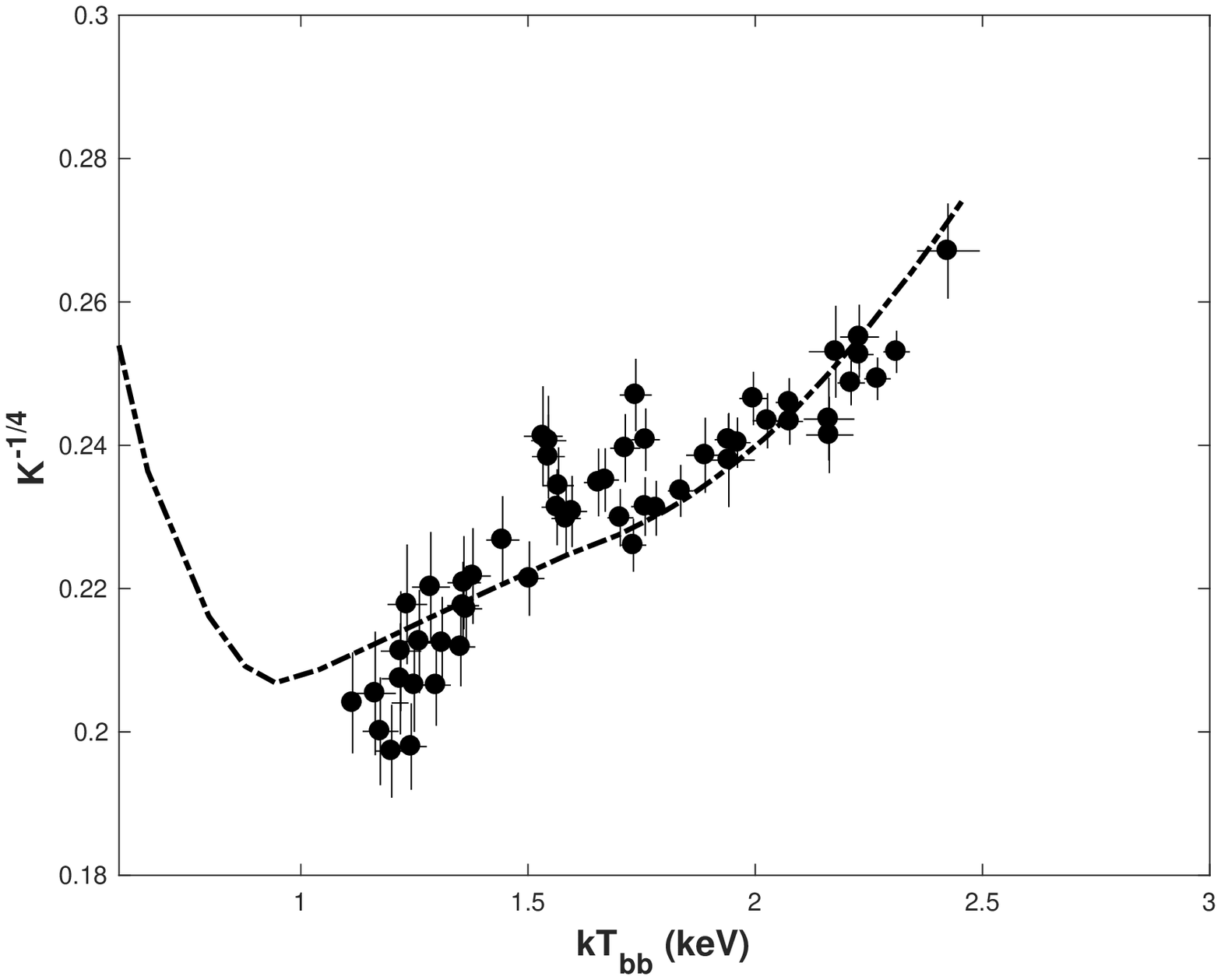}
\plottwo{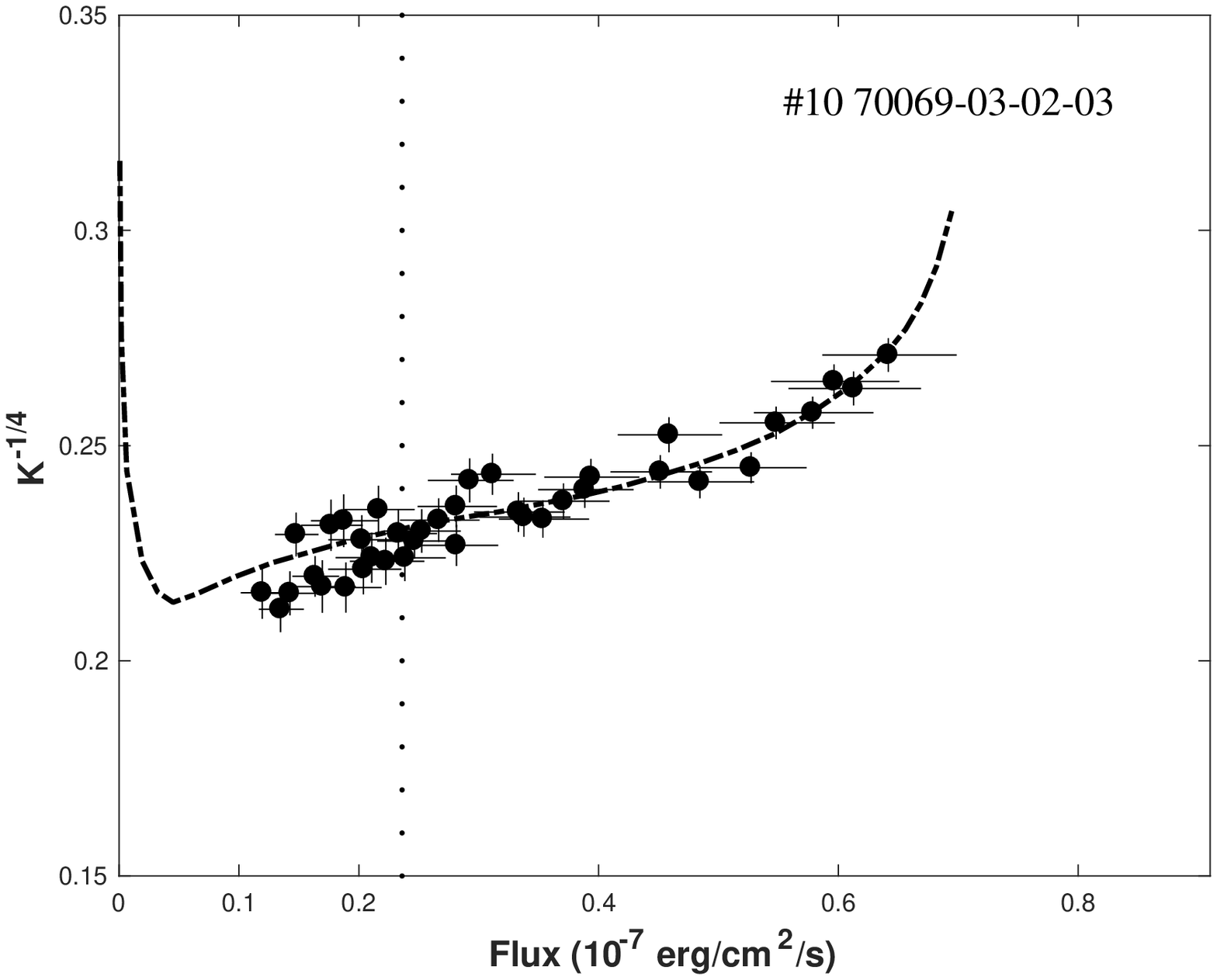}{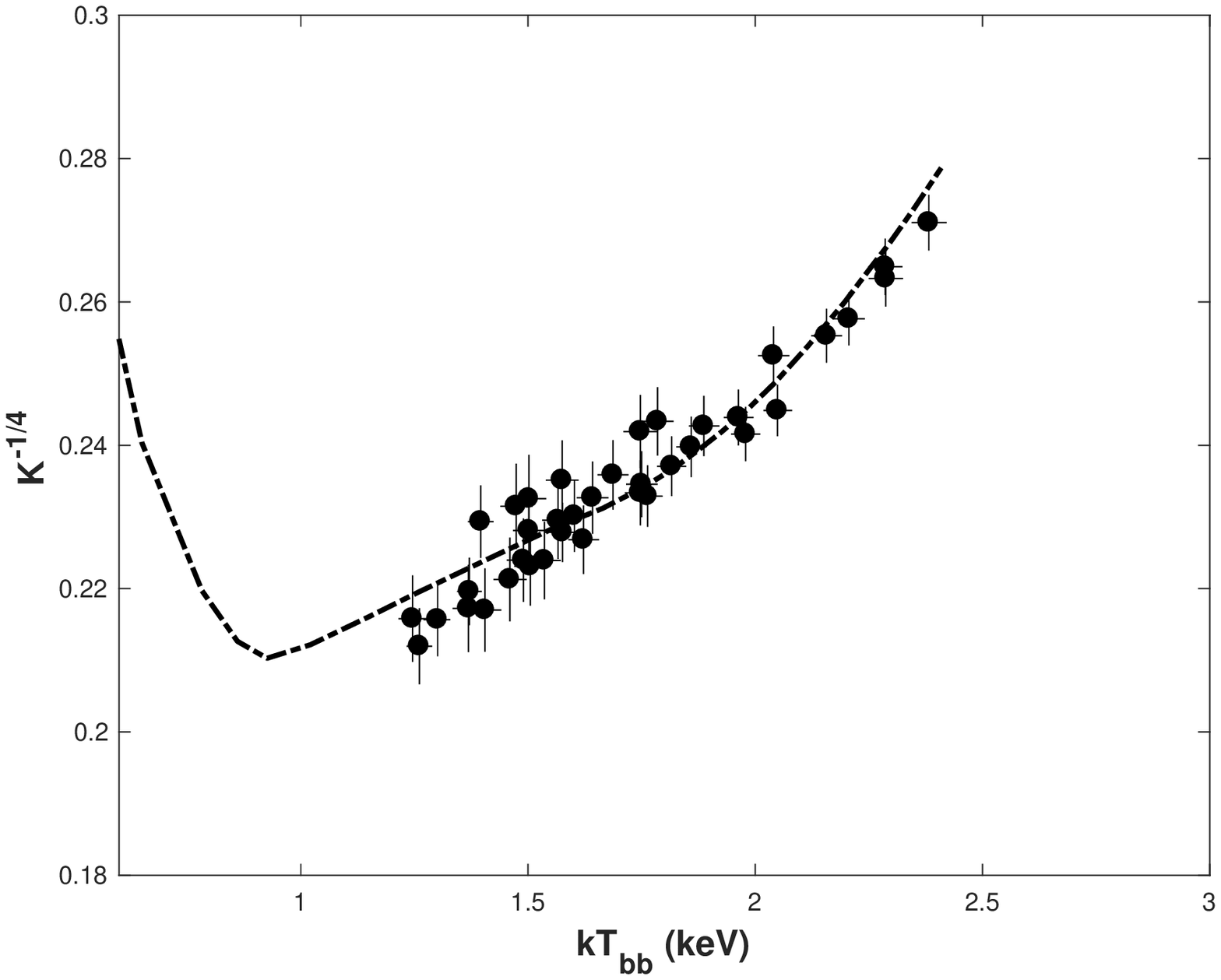}
\plottwo{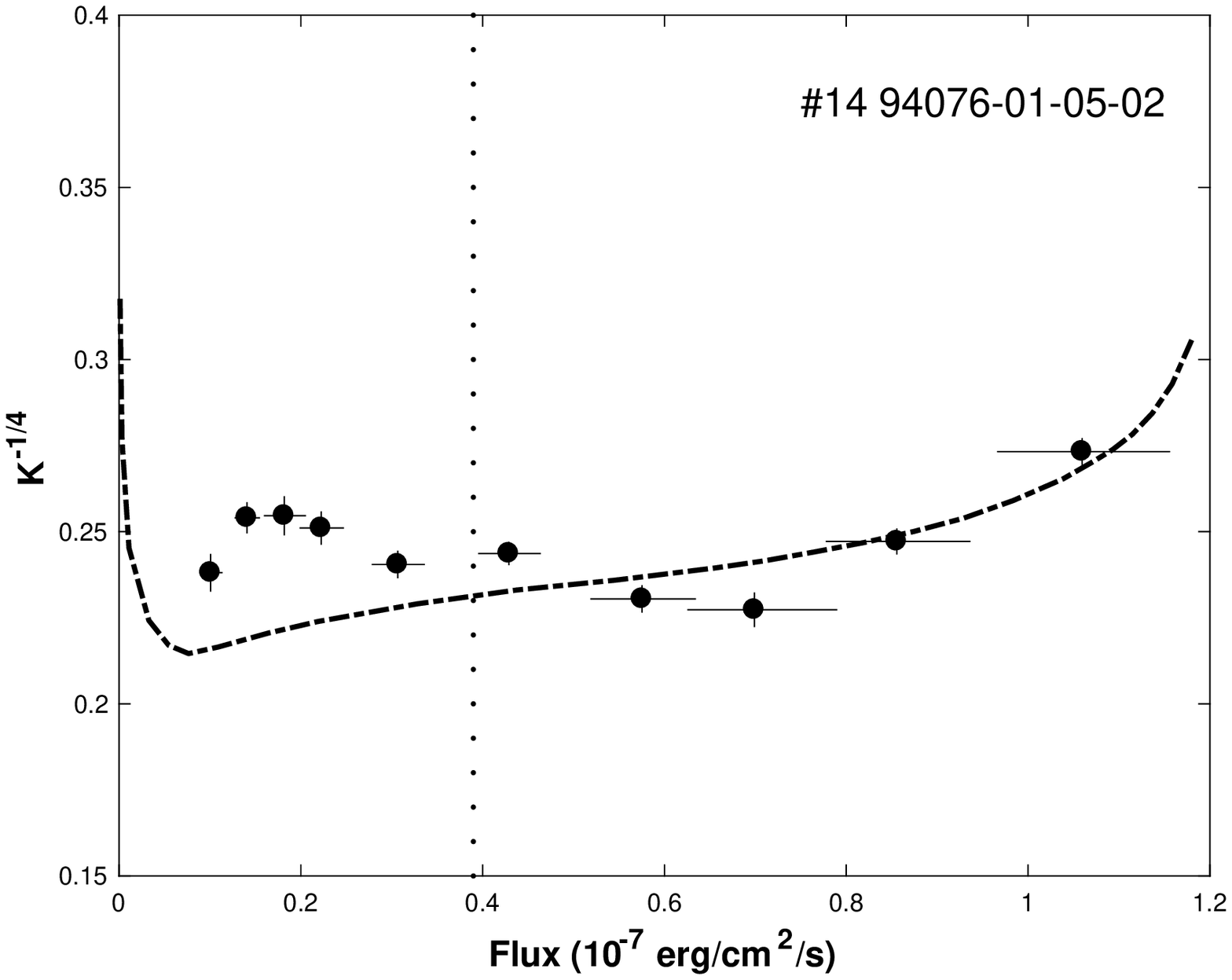}{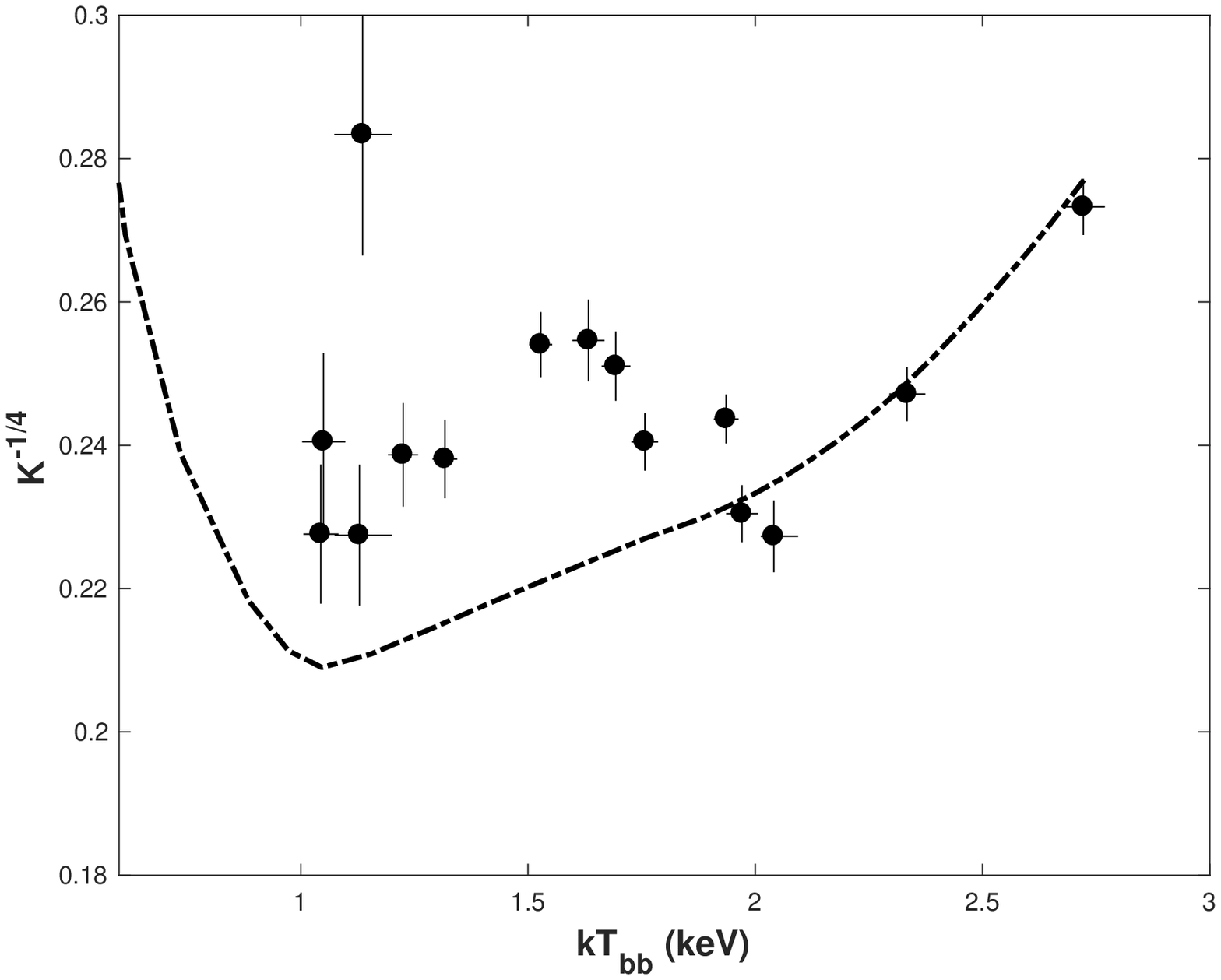}
\caption{The $F-K^{-1/4}$ (left panels) and  $kT_{\rm bb}-K^{-1/4}$ (right panels) tracks and their best fit model of the soft state PRE bursts (No. 2, 7, 10, 14, from top to down) in Aql X--1. The vertical dotted line in left panels presents the fitting truncated at $F_{\rm TD}/e$ \citep[see][for more details]{Poutanen14}.}\label{fig:cooling_soft}
\end{figure}

The best fit values and uncertainties of $A$ and $F_{\rm Edd}$ are obtained by the bootstrap method. The fit to the $kT-f_c$ relation is carried out in the similar way. We found, that 11 PRE bursts out of 13, fit the $kT_{\rm bb}-K^{-1/4}$ tracks better, with smaller $\chi_{\rm red}^2$, compared to the $F-K^{-1/4}$ trend. This confirms the finding that the data are best fit with the relation $kT_{\rm bb}-K^{-1/4}$  \citep[see also][]{Ozel15b}. The bursts with relative good fits are burst No 11, i.e., the hard state PRE burst, see Fig.~\ref{fig:cooling_hard} and  four PRE bursts in the soft state, see Fig.~\ref{fig:cooling_soft}. Three of them (Burst No. 2, 7, 10) have acceptable $\chi_{\rm red}^2$ values (both $\chi_{\rm red}^2<2$), however, the touchdown fluxes are in the range between $(6.2-7.3)\times 10^{-8}~\rm{erg\, cm^{-2}\, s^{-1}}$, which are only $\lesssim 60\%$ of the brightest bursts. If we use these samples to constrain the mass and radius of \object{Aql X--1}, it will underestimate the mass measurement.  For the hard state PRE burst (No. 11), the data can follow the trend of the model, but the $\chi_{\rm red}^2$ are larger. The theoretical model No. 8 from \citet{Suleimanov12} is not the only one to fit the data well. For the same abundance, the theoretical predictions are insensitively to the $\log g$ as explained in  \citet{Suleimanov12}. All fit results are listed in Table~\ref{table:burst}.

\begin{deluxetable*}{lcccccc}\centering
\tabletypesize{\scriptsize}
\tablecaption{Fitting to the $F-K^{-1/4}$ and $kT_{\rm bb}-K^{-1/4}$ relations.\label{table:fit}}
\tablewidth{0pt}
\tablehead{\colhead{Obs\_ID}&\colhead{Burst No.}   & \colhead{$F_{\rm Edd}$} & \colhead{A}      & $\chi_{\rm red}^2$({\it d.o.f})\tablenotemark{1}  & $\chi_{\rm red}^2$({\it d.o.f})\tablenotemark{2} &\\
\colhead{} &  & \colhead{$10^{-7}~{\rm erg\cdot cm^{-2}\cdot s^{-1}}$}    & \colhead{$(R(1+z)/D)^{-1/2}$}    &\colhead{}    &   
		  }
		\startdata
		20092-01-05-00&  {\#}1 &  $1.18\pm0.10$ & $0.163\pm0.002$    & 13.07(13) &6.09(13)& \\		
		20092-01-05-030 & {\#}2   &  $ 0.62\pm0.01$ & $0.168\pm0.001$  & 1.12(23)&1.37(23)& \\		
		20098-03-08-00&     {\#}3 &  $1.06\pm0.26$ &  $0.155\pm0.012$   & 12.86(11)&8.75(11)& \\		
		40047-03-02-00  &   {\#}4 & $ 1.10\pm0.15$ & $ 0.158\pm 0.009$  &16.97(16) &8.13(16)&\\		
		40047-03-06-00   &  {\#}5 & $ 1.18\pm0.01$ & $ 0.158\pm 0.002$  & 3.43(10)&1.69(10)& \\		
		40048-01-02-00  &  {\#}6  &- & - &-&-& \\		
		50049-02-13-01   &  {\#}7 &$0.84\pm0.01$ &$0.161\pm0.001$    &1.90(22) &1.38(22)&\\		
		60054-02-03-03   & {\#}8 &$0.64\pm0.16$  &$0.154\pm0.014$   &14.97(11)& 5.69(11)& \\		
		60429-01-06-00  &  {\#}9 &$0.80\pm0.31$ &$0.142\pm0.014$   & 15.74(16)&  7.58(16)&\\		
		70069-03-02-03 & {\#}10   &$0.64\pm0.01$ &$ 0.160\pm0.001$  &1.11(21)&0.65(21)& \\	
		92438-01-02-01  &   {\#}11&$0.85\pm0.04$& $0.133\pm 0.002$   & 10.34(28)&10(28)& \\ 		
		93405-01-03-07 &  {\#}12 &$0.83\pm0.06$ &$0.164\pm0.005$   &11.99(32)&6.14(32)& \\		
		94076-01-05-02 &   {\#}13 &$1.02\pm0.02$ &$ 0.160\pm0.002$ &8.45(25) & 7.85(25)&\\		
		96440-01-09-07 &  {\#}14 & $1.01\pm0.09$ & $0.163\pm0.003$    &4.63(3) &4.88(3)&
		\enddata
		\tablenotetext{1}{The reduced $\chi^2$ and degree of freedom from the fitting of $F-K^{-1/4}$.}
		\tablenotetext{2}{The reduced $\chi^2$ and degree of freedom from the fitting of $kT_{\rm bb}-K^{-1/4}$.}
	\end{deluxetable*}

  \subsubsection{Data-driven approach}
  
Based on the best data selection criteria introduced by \citet{Ozel15a, Ozel16}, a PRE burst, which can be used to determine the NS mass and radius, should satisfy at least two conditions. First, the touchdown flux should be the brightest ones among all samples, taking also the uncertainties into account. Second, the photospherical radius at the expansion phase should be evidently larger than the asymptotic value at the cooling tail.  The No. 11 PRE burst has larger mean value and standard derivation than other selected bursts which may be contaminated by the X-ray excess during the decay, see Fig.~\ref{fig:decay}.
For the bursts No. 9, the blackbody normalization increased during the cooling tail, so, these two bursts are not included in our samples. Totally, in our sample, the PRE bursts No. 1, 4, 12, 13 and 14 follow the two conditions. 



We only extracted the blackbody normalization in the range of $0.1\sim0.7~F_{\rm TD}$ \citep{Ozel16}, and the mean values are listed in Table~\ref{table:burst}.  The other selected PRE bursts have roughly the same blackbody normalizations, which are different from the noisy results extracted in the range between $(5\times 10^{-9}-F_{\rm TD})~{\rm erg\,cm^{-2}\,s^{-1}}$ from all X-ray bursts, i.e., PRE and non-PRE bursts \citep{Guver12a}.  Therefore, we obtain, the touchdown flux, apparent angular size $K$, the touchdown temperature $kT_{\rm TD}$ and their errors from the weighted mean and standard deviations of the selected samples, which are  $(1.06\pm0.10)\times10^{-7}~{\rm  erg\, cm^{-2}\, s^{-1}}$,  $279\pm30~\rm {(km/10~ kpc)^2}$  and $2.51\pm0.09~{\rm keV}$, respectively and are applied in our calculations. 
	


\section{Aql X--1 mass and radius}
  \label{sec:mass-radius}

The $M-R$ of Aql X--1 were independently constrained from the quiescent spectra observed by \textit{Chandra} and \textit{XMM-Newton}, as well as from the PRE bursts detected by \textit{RXTE}.

As the distance to Aql X--1 was not accurately measured, we divided the distance into $4\pm0.25~\rm{kpc}$, $4.5\pm0.25~\rm{kpc}$, $5\pm0.25~\rm{kpc}$, $5.5\pm0.25~\rm{kpc}$ and $6\pm0.25~\rm{kpc}$ with a box-car prior distribution for both methods, and simulated them separately.  From the quiescent spectra, we first run the MCMC simulations with the upper and lower limits of $N_{\rm H}$, which are $5.16\times10^{21}$ and $5.26\times10^{21}~\rm cm^{-2}$ respectively. The results are in well agreement with each other. We thus fix  $N_{\rm H}$ at $5.21\times10^{21}~{\rm cm^{-2}}$ to limit the computational time. The power-law index was tied for \textit{Chandra} and \textit{XMM-Newton} spectra and free to vary.  In addition, the NS atmosphere temperature, mass, radius, power-law normalization, distance were set as free parameters during the fit. For each of the 200 chains, the length was $2\times10^6$. The 20\% of the steps prior to the chains were burned. We also used the xspec\_emcee\footnote{https://github.com/jeremysanders/xspec\_emcee} program developed by Jeremy Sanders  to perform the MCMC simulations. The confidence regions of mass and radius are very similar to those obtained with the chain command in {\sc xspec}. The convergence of the MCMC was also verified, as the length of each chain is much larger ($> 600$) than the autocorrelation time of the mass and radius series and the chain of simulated parameters showed no significant trends or excursions. 
 	
\begin{figure*}\centering
\includegraphics[scale=0.5]{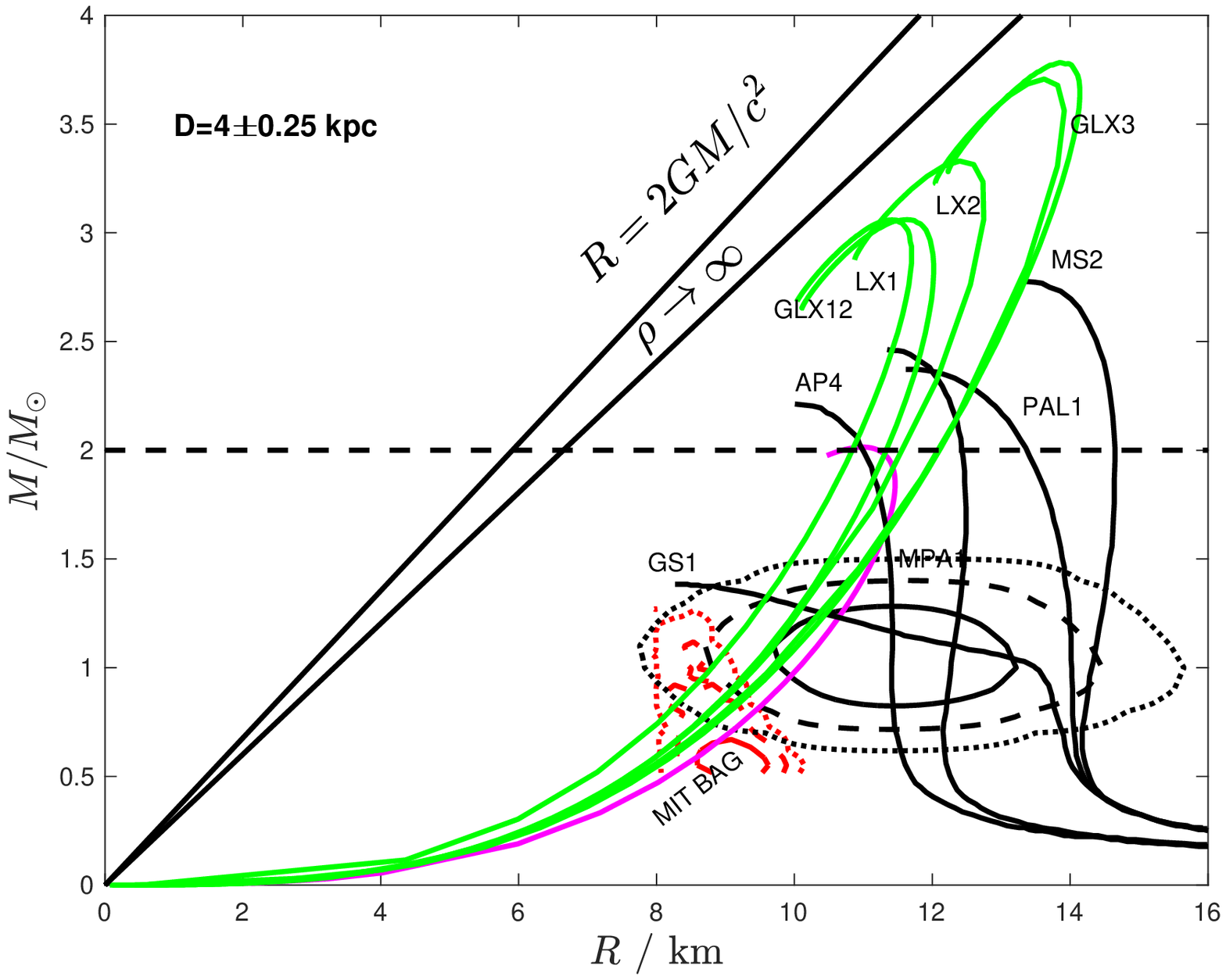}
\includegraphics[scale=0.5]{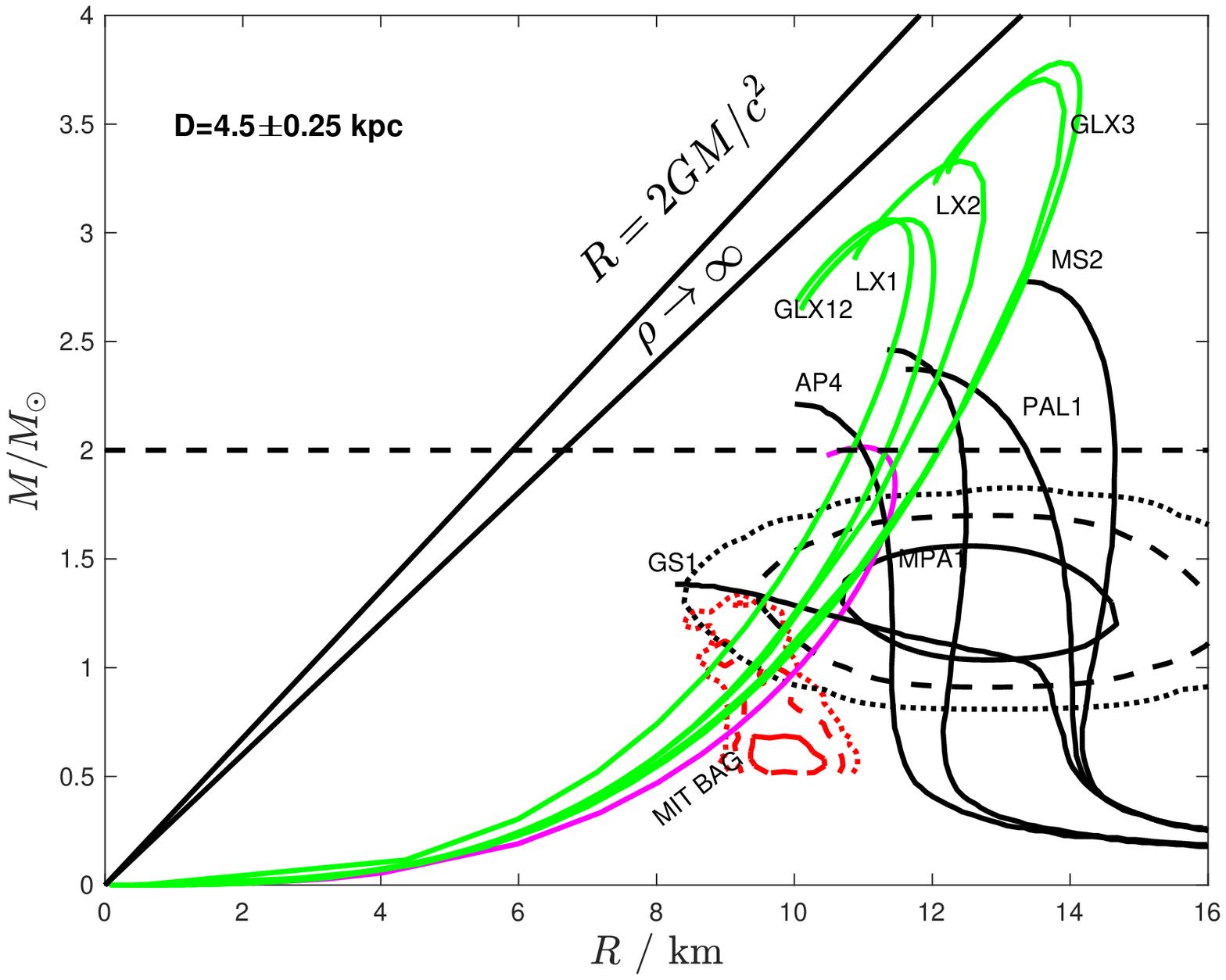}
\caption{The mass and radius of Aql X--1. The contours are obtained from PRE bursts (nearly horizontal black contours) and quiescent spectra (skewed red contours) respectively. The solid, dash and dotted contours represent the 1, 2, and 3$\sigma$ c.l. respectively. The distance is set as $4\pm0.25 ~ \rm{kpc}$ (\textit{left panel}) and $4.5\pm0.25~ \rm{kpc}$ (\textit{right panel}). In both panels, the dashed line labels two observed near $2M_\odot$ NSs. The two black straight lines show the constrains from the general relatively (GR) and the central density limit, respectively. Theoretical mass-radius relations were predicted for several NS EoS models, which are marked as GS1 \citep{Glendenning99}, AP4 \citep{AP97}, MPA1 \citep{MPA87}, PAL1 \citep{PAL88}, MS2 \citep{MS96}, GLX123 \citep{Guo13}, LX12 \citep{Lai09,Lai13}. The purple dash-dotted line represents the bare strange stars obtained from MIT bag model EoS with the bag constant $57~{\rm MeV/fm^3}$. First five EoSs are gravity bound, while the rest of them are self bound on surface. }\label{fig:eos1}
\end{figure*}

\begin{figure*}\centering
\epsscale{1.05}
\plottwo{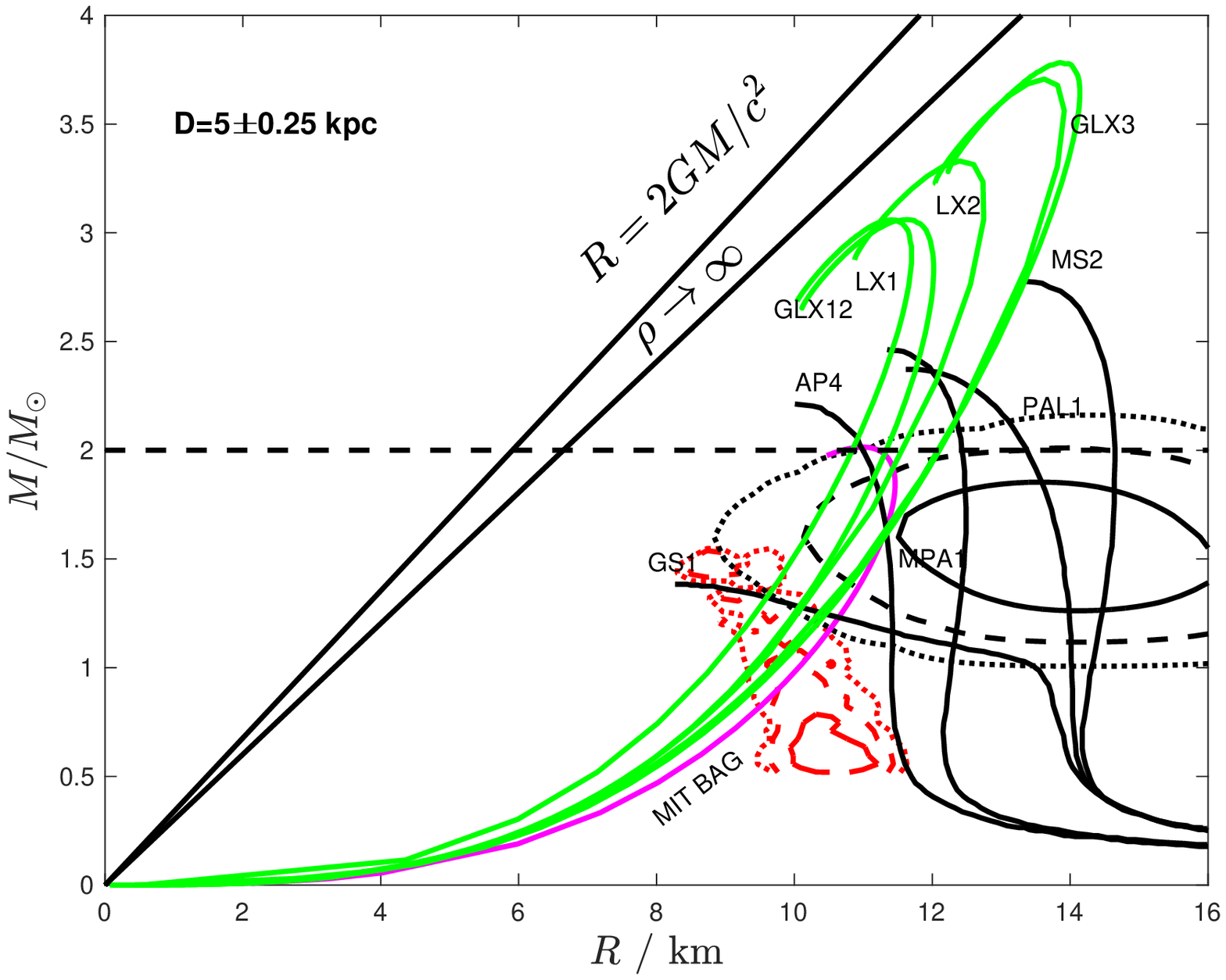}{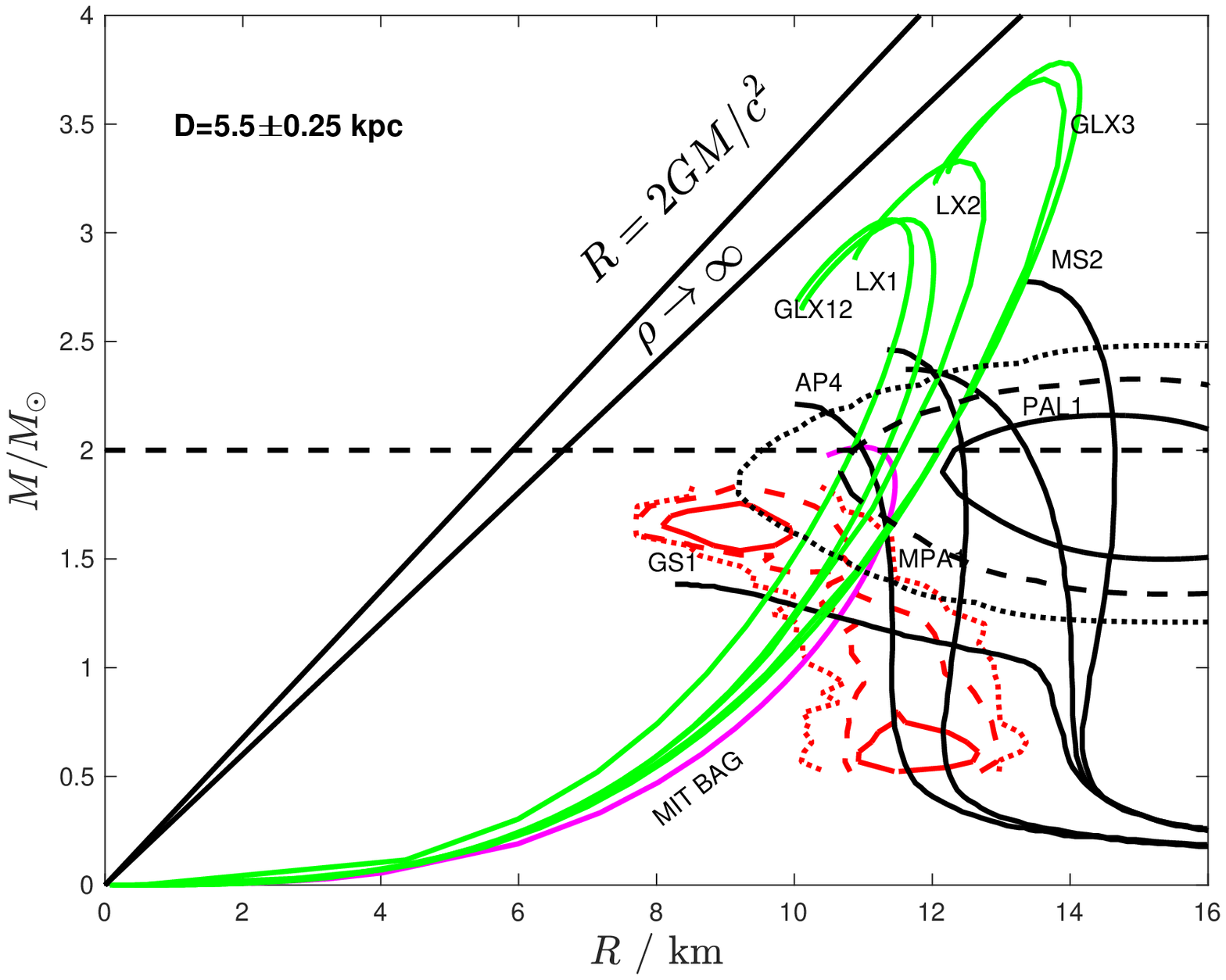}
\caption{Similar as Figure~\ref{fig:eos1}. But the distance is $5\pm0.25 ~ \rm{kpc}$ (\textit{left panel}) and $5.5\pm0.25 ~\rm{kpc}$ (\textit{right panel}).  }\label{fig:eos2}
\end{figure*}

From PRE bursts, we adopted the Bayesian approach to measure the mass and radius of the NS. In Eq.~(\ref{equ:prob}), the corrected touchdown flux and apparent angular size were applied based on Eq.~(\ref{equ:flux_cor}) and (\ref{equ:area_cor}).  We chose prior Gaussian distributions for  $kT_{\rm TD}$, $K$ and $F_{\rm TD}$, and prior flat distributions for distance, the hydrogen mass fraction, as well as the color correction factor, which are expressed as $D\sim U[D-{\rm d}D,~D+{\rm d}D]$, $X\sim U[0.3,~1]$, $f_{\rm c}\sim U[1.35,~1.45]$, respectively.  The NS mass and radius obtained from PRE burst method are shown as nearly black horizontal contours in Fig.~\ref{fig:eos1} and \ref{fig:eos2}. We only displayed the results for distances as high as $5.5\pm0.25~\rm{kpc}$, as no overlapping region exists combining these results with the qLMXB method at higher distances. 

The mass and radius of \object{Aql X--1} are shown in Fig.~\ref{fig:eos1} and \ref{fig:eos2}. The horizontal and skewed contours are obtained by using the results from  PRE bursts and quiescent spectra, respectively. The distance and its uncertainty are taken into account in the simulations. In each case, the uncertainty is $0.25~ \rm{kpc}$, while the distance spans the range $4-5.75~ {\rm kpc}$. From these results, we found, that the overlapped $M-R$ confident regions are always consistent with the strange matter EoSs (e.g., quark star and quark-cluster star, \citealt{Lai09,Lai13,Guo13}), and only the conventional neutron star EoS \citep{AP97} for the distance of $5.25-5.75~{\rm kpc}$. 
In addition, no confident regions are overlapped between quiescent spectra and PRE bursts when the distance is above  $5.75~ {\rm kpc}$. So, we can roughly estimate a distance range of $4-5.75~{\rm kpc}$. 
	
\section{Discussions}
\label{sec:discussions}
\subsection{Non-uniform emission due to residual accretion?}
The power-law component in quiescent spectra shows that the residual accretion may occur during the quiescent state. In our fitting procedure, we assume a uniform temperature  on the NS surface. If the accreted matter has a non-uniform distribution in a long time scale, our assumption could be questionable. However,  assuming that the timescale for the accreted matter to diffuse over the surface is $\tau_{{\rm diff}}$ in a random walk approximation, one has $R \sim N^{1/2} r_L$, where $R$ is the stellar radius, $r_L \sim 3\times10^{-2} (B/{10^8~{\rm G}})^{-1}~ {\rm cm}$ the Larmor radius, and $N = \tau_{\rm diff} / \tau_c$  the number of collision within the collision timescale  $\tau_c \sim 10^{-14}~{\rm s}$ \citep[e.g.,][]{Goldston95, Xu14}.  If $B \sim 10^8~{\rm G}$ and the density of the heated matter is in the order of $10^2 ~{\rm g/cm^3}$, we have $\tau_{\rm diff} \sim10 ~{\rm s}$, which is much shorter than the characteristic time of heat conduction and cooling. The accreted matter is thus likely to diffuse quickly and nearly heat almost the entire NS surface if the stellar magnetic field is weak ($B \sim 10^8~\rm{G}$  on surface). Furthermore, the pulse fraction was only about 2\% even in continuous accretion process \citep{Casella08}, which implies a negligible inhomogeneous temperature distribution. We thus conclude that the NS surface is uniformly emitting, even if  the residual accretion occurs during quiescence.  

\subsection{The possible reasons of $M-R$ overlapped at $1\sigma$ c.l. do not exist}
We notice that there are no overlapping $M-R$ relation between the PRE bursts and qLMXB results at $1\sigma$  c.l. This can be explained in at least  three scenarios. First, we assumed a pure hydrogen NS  atmosphere in the {\sc nsatmos} model. However, the atmosphere on the surface of Aql X--1 should be composed by a mix of hydrogen and helium, as its companion is a main sequence star. Second, we corrected the fast rotation effects of Aql X--1 in the PRE burst method, but these theoretical calculations are still not taken into account in quiescent NS atmosphere models. The combination of  fast rotation and hydrogen and helium mixed NS atmosphere in a refined spectral model could help us in understanding better the quiescent spectra of \object{Aql X--1}.  Third, we assume that the photosphere expands in spherical symmetry. However, the diversity of the mean cooling area implies that the photosphere expansion might be asymmetric, which bias the NS mass and radius measurements.

\section{Summary}
\label{sec:summary}
For the first time the mass and radius of \object{Aql X--1} were constrained by PRE bursts and quiescent spectra simultaneously. As these two methods are completely independent, the NS mass and radius can give better constrains. 


14 PRE bursts were observed in \object{Aql X--1}, and only one of them during the source hard state. \citet{Poutanen14} suggested that only PRE bursts in the hard state can be used to determine the NS mass and radius, as the accretion rate is relatively low in the hard state and the accretion disk is not expected to produce obvious effects on the cooling tracks. Indeed, the cooling track of \object{Aql X--1} in the hard state follows the prediction of \citet{Suleimanov11,Suleimanov12} with large $\chi_{\rm red}^2$. We found, in any case,  that the three soft state PRE bursts followed the theoretical  prediction of the $F-K^{-1/4}$ and $kT_{\rm bb}-K^{-1/4}$ relations, but the touchdown fluxes are apparently smaller than the brightest ones. In addition, \citet{Ozel15a} proposed that the rapid evolution of the color correction factor could be missed at touchdown due to limitations of \textit{RXTE}. 


From the quiescent data observed by \textit{Chandra} and \textit{XMM-Newton}, we fitted the spectra with an absorbed hydrogen atmosphere emission component plus an hard power-law component. We used the Goodman-Weare algorithm of MCMC to simulate  the NS mass and radius for various prior distance distributions. The NS mass and radius in \object{Aql X--1} are shown in Fig.~\ref{fig:eos1} and \ref{fig:eos2}. Ten EoSs were also plotted to illustrate the constraints of \object{Aql X--1}. Once the results from the PRE bursts observed by \textit{RXTE} are combined with the quiescent spectra observed by \textit{Chandra} and \textit{XMM-Newton}, the mass and radius of \object{Aql X--1} are found to compatible with the strange matter EoSs \citep{Lai09,Lai13,Guo13} and the conventional neutron star EoS \citep{AP97}.  Moreover, we also concluded that the distance to \object{Aql X--1} should be in the range of $4.0-5.75~{\rm kpc}$ because no overlapped $M-R$ confidence region exists when higher distances are considered.	

The EoSs of a compact star could be strictly tested by very high mass NSs, very low mass NSs \citep{Li15}, as well as accurately measurements of NS mass and radius. In this work, we applied simultaneously two well established methods  for \object{Aql X--1}, which could effectively reduce the mass and radius uncertainties  (see Fig.~\ref{fig:eos1} and ~\ref{fig:eos2}). Precious distance measurements from optical observations (such as Thirty-Meter Telescope) could help in obtaining tighter constraints on the NS EoSs. Better constraints on the NS EoSs will be obtained in the future by expecting the advanced capabilities of eXTP \citep{Zhang16} which will be able to measure simultaneously the NS pulse profile with high accuracy, its quiescent spectrum with larger signal to noise ratio and collect  both PRE bursts and gravitational redshift measurements.

\section*{Acknowledgments}
We thank the anonymous referee for the comments and suggestions which significantly improve our manuscript. Z. Li is supported by the Swiss Government Excellence Scholarships. Z. Li thanks the International Space Science Institute for the hospitality during his visiting, L. Zampieri for providing his Xspec table model generously, and E. Bozzo for carefully reading this manuscript.  This work is supported by the National Natural Science Foundation of China (11225314, 11173024), Hunan Provincial NSF 2017JJ3310, the Strategic Priority Research Program on Space Science of the Chinese Academy of Sciences (XDA04010300). This research has made use of data obtained from the High Energy Astrophysics Science Archive Research Center (HEASARC), provided by NASA's Goddard Space Flight Center. The FAST FELLOWSHIP is supported by the Special Funding for Advanced Users, budgeted and administrated by Center for Astronomical Mega-Science, CAS.

\end{document}